\documentclass[10pt]{article}

\usepackage{ragged2e}       
\usepackage{soul}           
\usepackage{xcolor}
\usepackage{graphicx}
\usepackage{hyperref}
\usepackage{amsmath}
\usepackage{subcaption}
\usepackage{amssymb}
\usepackage{booktabs}
\usepackage{fullpage}
\usepackage[section]{placeins}
\usepackage{authblk}
\usepackage{tabularx}
\usepackage{enumerate}

\definecolor{my_purple}{RGB}{255,0,255}
\graphicspath{{SphinctFigures_3/}}

\definecolor{REDCOLOR2}{RGB}{0,0,0}

\title{\textbf{\Large A mechanics-based perspective on the function of human sphincters during functional luminal imaging probe manometry}}

\author[1]{\normalsize Guy Elisha}
\author[2]{\normalsize Sourav Halder}
\author[3]{\normalsize Dustin A. Carlson}
\author[3]{\normalsize Wenjun Kou}
\author[3]{\normalsize Peter J. Kahrilas}
\author[3]{\normalsize John E. Pandolfino}
\author[1,2]{\normalsize Neelesh A. Patankar\thanks{Corresponding author: N.~A.~Patankar (\texttt{n-patankar@northwestern.edu})}}

\affil[1]{Department of Mechanical Engineering, McCormick School of Engineering, \newline
Northwestern University Technological Institute, 2145 Sheridan Road, Evanston, IL 60201 \vspace{1ex}}
\affil[2]{Theoretical and Applied Mechanics Program, McCormick School of Engineering,\newline
Northwestern University Technological Institute, 2145 Sheridan Road, Evanston, IL 60201 \vspace{1ex}}
\affil[3]{Division of Gastroenterology and Hepatology, Feinberg School of Medicine, \newline
Northwestern University, 676 North St. Clair Street, Arkes Suite 2330, Chicago, IL 60611 \vspace{1ex}}

\date{}

\begin{document}

\maketitle 

\begin{abstract}
Functional luminal imaging probe (FLIP) is used to measure cross-sectional area (CSA) and pressure at sphincters. It consists of a catheter surrounded by a fluid filled cylindrical bag, closed on both ends. Plotting the pressure-CSA hysteresis of a sphincter during a contraction cycle, which is available through FLIP testing, offers information on its functionality, and can provide diagnostic insights. However, limited work has been done to explain the mechanics of these pressure-CSA loops. This work presents a consolidated picture of pressure-CSA loops of different sphincters. Clinical data reveal that although sphincters have a similar purpose (controlling the flow of liquids and solids by opening and closing), two different pressure-CSA loop patterns emerge: negative slope loop (NSL) and positive slope loop (PSL). We show that the loop type is the result of an interplay between (or lack thereof) two mechanical modes: (i) neurogenic mediated relaxation of the sphincter muscle and (ii) muscle contraction proximal to the sphincter which causes mechanical distention. We conclude that sphincters which only function through mechanism (i) exhibition NSL whereas sphincters which open as a result of both (i) and (ii) display a PSL. This work provides a fundamental mechanical understanding of human sphincters. This can be used to identify normal and abnormal phenotypes for the different sphincters and help in creating physiomarkers based on work calculation. 
\end{abstract}


\section{Introduction}\label{Introduction}

A sphincter is a tonically contracted circular muscle which controls the flow of liquids and solids. For example, the lower esophageal sphincter (LES) monitors the entrance of swallowed matter from the esophagus into the stomach while preventing retrograde flow from the stomach into the esophagus (acid reflux) \cite{Hershcovici2011}. The human body includes many sphincters, with main examples listed in table \ref{table:SphincterList}. Dysfunction of sphincters is the root cause of many physiological disorders, especially along the GI track \cite{Gregersen2018}. In the case of the LES, dysfunction can be either excessive compliance causing retrograde acid flow into the esophagus (such as in gastroesophageal reflux disease \cite{Pandolfino2003,Kahrilas2008}) or absence of LES relaxation preventing healthy opening of the sphincter and emptying of the esophagus (such as in achalasia \cite{Boeckxstaens2014,Eckardt2009}). Hence, it is important to understand the mechanical function of the sphincters, and particularly, their opening and closing mechanisms \cite{Gregersen2018}.


\begin{table}[!htb]
\fontsize{8}{10}\selectfont
\centering
   \caption{List of sphincters in the human body}
    \label{crouch}
   \begin{tabular}{p{0.15\textwidth}  p{0.75\textwidth}  }
        \toprule
\textbf{Sphincter}      
& \textbf{Description} \\\midrule
Upper esophageal sphincter (UES)
& A C-shaped striated muscle sphincter attached to the lateral aspects of the cricoid cartilage \cite{Kahrilas2022}. It does not have a resting tone, but rather responds to a vast array of reflexive inputs \cite{Kahrilas2022}. Its main function is to monitor the flow coming into and out of the esophagus. During respiration, the UES is to remain occluded to prevent air from entering the esophagus and the esophagus content from flowing back into the mouth. During swallowing, the UES opens to allow bolus transport from the mouth into the esophagus \cite{Sivarao2000,patel2018endoscopic}.\\\hline
Lower esophageal sphincter (LES)       
& Tonically contracted composite of different muscles located at the distal end of the esophagus. The LES moderates the entrance of swallowed material coming from the mouth through the esophagus into the stomach while preventing gastroesophageal reflux \cite{Hershcovici2011,preiksaitis1994nitric,Mittal1997}.\\\hline
Pyloric sphincter (PS)        
& Circular muscle layer connecting between the pylorus and the duodenum which controls the flow of stomach discharge into the duodenum \cite{moore2018clinically,shafik2006mechanism}. Its function and opening depends on the phase: digestive or inter-digestive.\\\hline
Sphincter of Oddi (SoO) &
Tonically contracted at rest, located at the end of the ampulla (between ampulla and small intestine). The SoO regulates the flow of bile and pancreatic juice into the small intestine and prevents reflux of small intestine content into the ampulla \cite{moore2018clinically,woods2005sphincter}. \\\hline
Internal anal sphincter (IAS)&
Smooth muscle sphincter located around the anal canal \cite{moore2018clinically}. It is always in a state of tonic contraction which is maintained by sympathetic fibers to prevent leakage \cite{moore2018clinically}. The sphincter relaxes involuntary when sufficient fecal material accumulates in the canal \cite{moore2018clinically,meunier1977control}. This does not imply defecating since the EAS also plays a role \cite{moore2018clinically}.\\\hline
External anal sphincter (EAS)&
Striated muscle sphincter located around the IAS \cite{moore2018clinically}. It is always in a state of tonic contraction to keep the anal canal shut and prevent leakage. Once a voluntary signal is sent to relax the sphincter, it opens and allows the passage of feces out of the body \cite{moore2018clinically}. The EAS can be further contracted (occluded) voluntarily \cite{moore2018clinically}.\\\hline 
Internal urethral sphincter (IUS)&
Smooth muscle fibers located at the junction of the urethra with the urinary bladder \cite{jung2012clinical}. Its main role, together with the EUS, is to control the flow of urine \cite{moore2018clinically}. The IUS is an involuntary sphincter that is tonically contracted at rest and relaxes during micturition \cite{jung2012clinical,de1976reflex}. The relaxation is controlled by the nervous system \cite{chancellor2004neurophysiology}. For males, the IUS is also responsible for preventing retrograde flow of semen into the bladder during ejaculation \cite{moore2018clinically,sam2020anatomy}.\\\hline 
External urethral sphincter – female (EUSF)&
Fibrous muscle which surrounds the urethra in the middle third of its length \cite{jung2012clinical}. It is located between the vaginal orifice and the clitoris \cite{sam2020anatomy}. The EUSF is a voluntary sphincter which is tonically contracted to prevent the leakage of urine. Once it relaxes and opens, it allows the flow of urine out of the body \cite{moore2018clinically,yucel2004anatomical}.\\\hline   
External urethral sphincter – male (EUSM)&
Circular muscle fibers located between the pudendal canals and below the pelvic diaphragm \cite{sam2020anatomy}. The sphincter muscles are tonically contracted to prevent leakage of urine, and relax voluntary to allow voiding \cite{moore2018clinically}. \\
        \bottomrule
    \end{tabular}
    \label{table:SphincterList}
\end{table}

\begin{figure*}[!htb]
    \centering{{\includegraphics[trim=0 0 0 0 ,clip,width=0.9\textwidth]{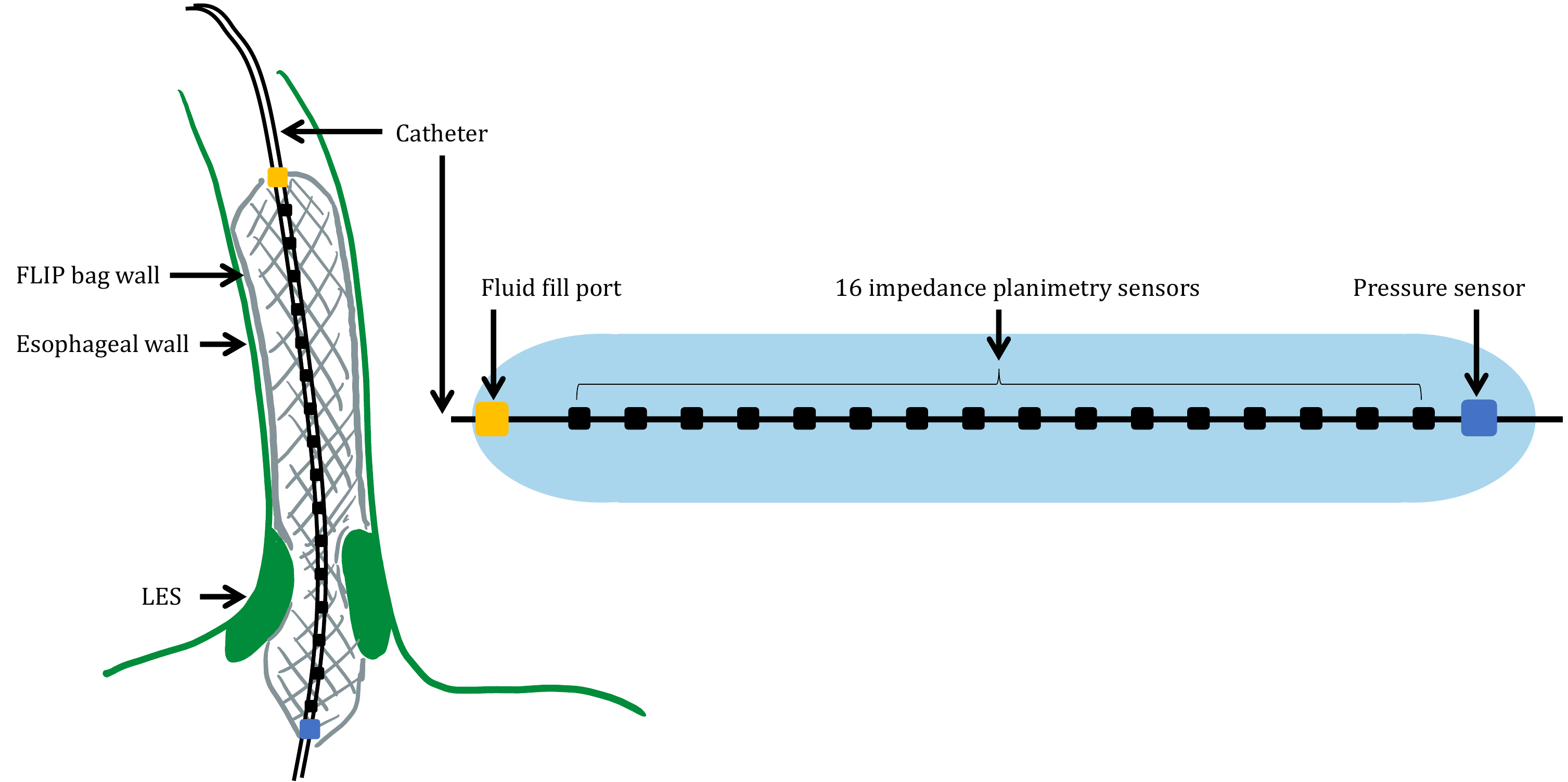}}}
    \caption{Diagram showing the FLIP bag and catheter assembly placed within the esophagus, bypassing the LES. The bag used in this study is 16 cm long and 22 mm diameter. The catheter measures CSA at 16 location and one pressure measurement at each time instance.}
    \label{fig:FLIP_bag}
\end{figure*}

As noted by \cite{Gregersen2018}, a functional luminal imaging probe (FLIP) has increasingly become one of the more useful tools to determine the shape of a sphincter. The FLIP consists of a catheter and a bag which is placed inside a channel (such as  the anal canal), bypassing the sphincter. The FLIP measures cross-sectional area (CSA) and pressure in the bag as a function of time, as the walls of the sphincter respond to the filling of the bag \cite{Hirano2017,Lottrup2015}. The FLIP bag length and volume depends on the experimental set up. For example, in the FLIP study on the anal sphincter (AS) by \cite{Zifan2019}, the bag length was 30 mm, and the maximum bag volume was 90 ml. In a FLIP study by \cite{Carlson2015} on the esophagus, the FLIP bag length was 16 cm, and the maximum bag volume was 60 ml. Figure \ref{fig:FLIP_bag} presents an illustration of a FLIP device used in the esophagus.

The procedures themselves also differ based on the physiology of the sphincter. For instance, in the case of the esophagus and the LES, the contraction and relaxation during a FLIP testing is mainly involuntary. As the volume of the bag increases, it triggers distention induced peristalsis, which varies the CSA and bag pressure \cite{Carlson2015}. On the other hand, Zifan et al.'s~\cite{Zifan2019} study captured voluntary contraction of the AS (active squeezing of the sphincter muscle).

The FLIP has become increasingly popular since it allows distensibility testing; the sphincter's resistance to distention. This evaluation can be easily done using FLIP over other measuring devices. Hence, it is known to be the superior approach in evaluating the condition of a sphincter \cite{Gregersen2018,Gregersen2016}. 

Borrowed from cardiovascular assessment, one approach taken to study the mechanical function of sphincters is by tracking its pressure and CSA during an opening and closing cycle \cite{Gregersen2018}. Plotting the pressure-CSA hysteresis of a sphincter offers information on its functionality and can provide diagnostic insights. Moreover, it unveils the way in which work is done and expended during a contraction cycle \cite{Gregersen2018,Elisha2021Loop}. Producing a pressure area loop at the sphincter was made possible by FLIP testing. 


Our goal is to provide a framework in which we can investigate the function of different sphincters. In this work, we seek to provide a mechanics-based perspective on the function of human sphincters during FLIP manometry. While many studies have used FLIP to evaluate sphincter functionality, to our knowledge, no work has been done to create a common understanding of FLIP data across different sphincters. In addition, limited work has been done to explain the mechanics of the pressure-CSA loops at the different sphincters. Thus, in this study, we aim to find pressure-CSA hysteresis characteristics for each sphincter and identify the mechanisms that create such patterns. We do this through a literature survey of the pressure-CSA loops available and introduce additional clinical data on pressure-CSA loops at the LES. Furthermore, we explain the sphincter opening and closing pattern by reproducing them using simulations and looking at the work done by the sphincter wall throughout an opening and closing cycle. 

\section{Analysis of Literature Data}\label{LitAnalysis}

\begin{figure*}

    \centering
    \begin{subfigure}[b]{0.3\textwidth}
        \centering
        \includegraphics[trim=30 170 60 175,clip,width=\textwidth]{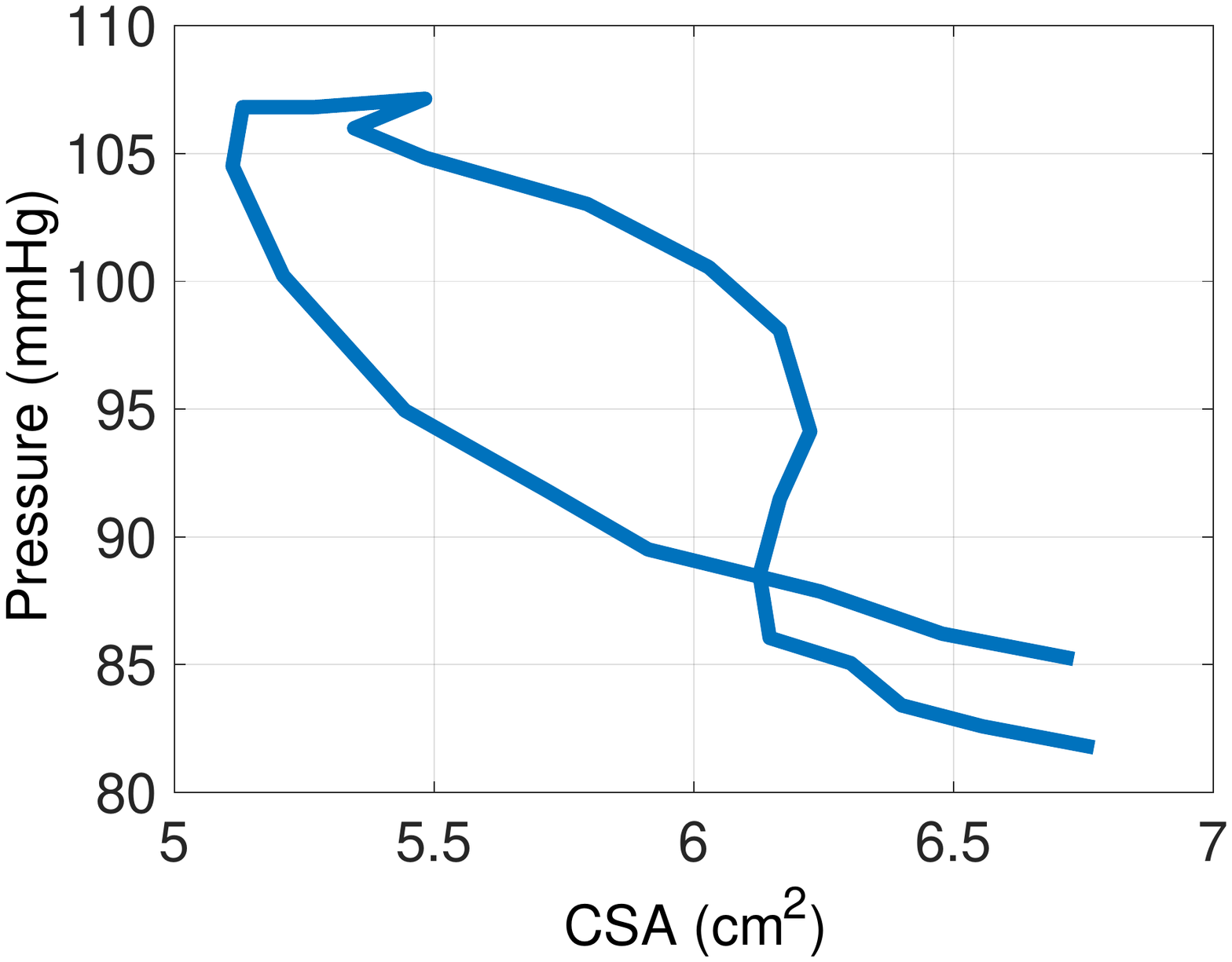}
        \caption{AS by \cite{Zifan2019}}
        \label{fig:AS_lit}
    \end{subfigure}
    \ 
    \begin{subfigure}[b]{0.3\textwidth}  
        \centering 
        \includegraphics[trim=30 170 60 175,clip,width=\textwidth]{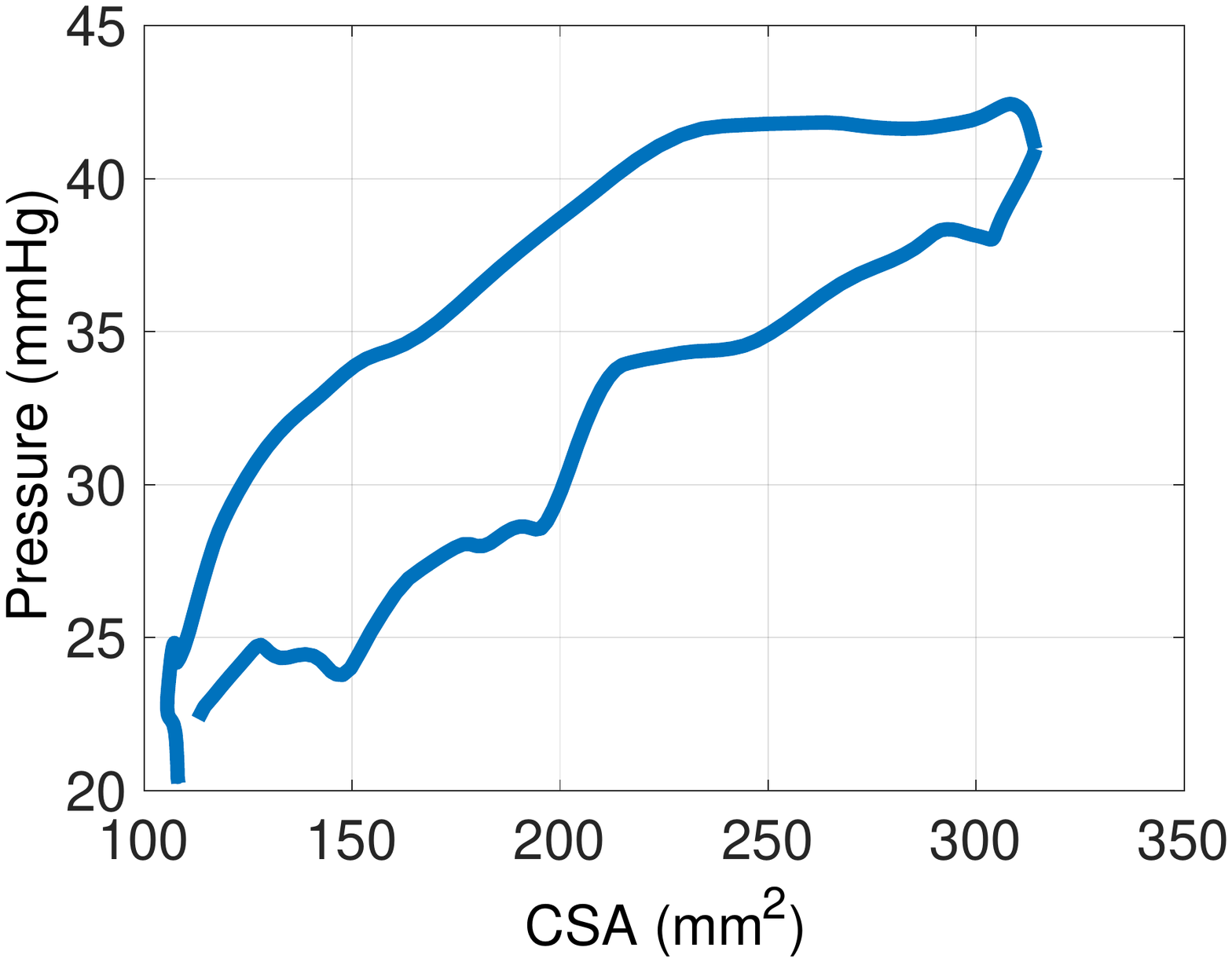}
        \caption{LES}
        \label{fig:LES_lit}
    \end{subfigure}
    \ 
    \begin{subfigure}[b]{0.3\textwidth}   
        \centering 
        \includegraphics[trim=30 170 60 175,clip,width=\textwidth]{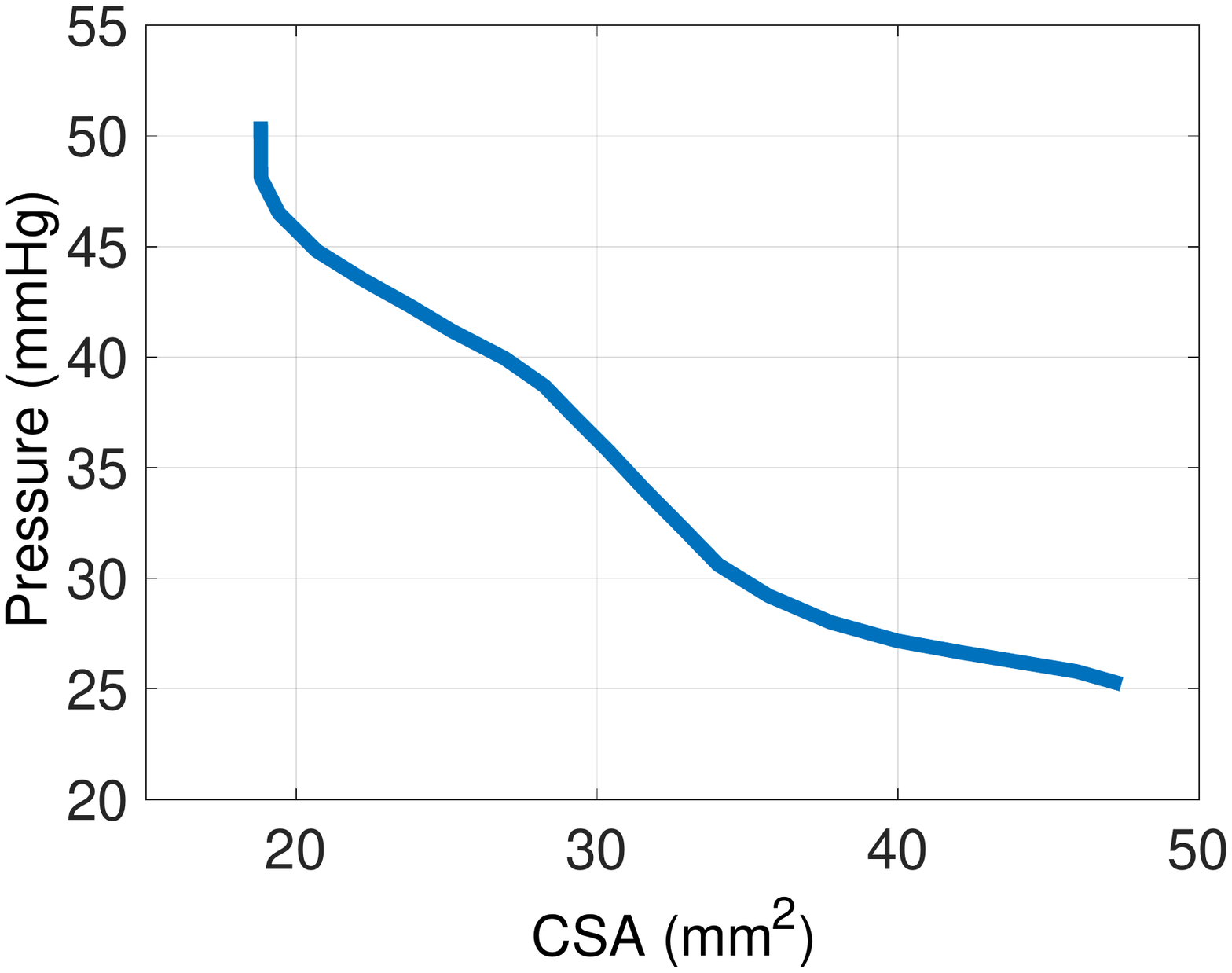}
        \caption{SoO curve by \cite{kunwald2010new}}
        \label{fig:SO_lit}
    \end{subfigure}
    \caption{Pressure-CSA plots obtained by FLIP experiments reported in literature.}
    \label{fig:litLoops}
\end{figure*}

Table \ref{table:LitList} presents a summary of studies which used FLIP or FLIP-like devices to investigate and report pressure and CSA values at different sphincters. Out of these studies, only one by Zifan et al.~\cite{Zifan2019}, plotted the pressure-CSA loops at the sphincter during a single contraction cycle (figure \ref{fig:AS_lit}). In addition, given our clinical experiment, detailed in section \ref{Method}, we managed to obtain such loops for the LES (figure \ref{fig:LES_lit}). The study by Kunwald et al.~\cite{kunwald2010new} on the Sphincter of Oddi (SoO) presents curves for pressure vs. CSA during a pressure ramp, as shown in figure \ref{fig:SO_lit}, but not a complete loop. To the best of our knowledge, no clinical pressure-CSA loops, during a single contraction cycle, have been plotted for the other sphincters listed in table \ref{table:SphincterList}.

Note that many studies have reported pressure-CSA loops at both female urethral sphincter (FUS), and male urethral sphincter (MUS) using a FLIP-like device of a catheter surrounded by a cylindrical shaped bag \cite{klarskov2007urethral,aagaard2012urethral,khayyami2016}. However, these loops do not represent a single contraction cycle. Each point on the loop represents pressure and CSA at a different bag volume. We did not find any studies capturing pressure and CSA behavior of the urethral sphincter (US) with a bag of constant volume.



\begin{table}[!htb]
\fontsize{8}{10}\selectfont
\centering
   \caption{List of literature of FLIP and FLIP-like studies at the different sphincters}
    \label{crouch}
    \begin{tabular}{  p{0.1\textwidth} p{0.1\textwidth}  p{0.1\textwidth} p{0.1\textwidth} p{0.4\textwidth} }
        \toprule
\textbf{Reference}      
& \textbf{Placement}
& \textbf{Bag geometry}   
& \textbf{Subjects (n)}   
& \textbf{Description} \\\midrule

Regan et al. 2013 \cite{regan2013new}    
& UES
&25-mm diameter
10-cm long
& 11 (controls)                           
& The FLIP bag was placed across the UES and the bag was filled. The CSA and pressure in the bag were measured during wet and dry swallows at each bag volume (5,10,15,20 mL). \\\hline

Kwiatek et al. 2010 \cite{Kwiatek2010}    
& LES
&25-mm diameter
6.4-cm long
& 20 (controls)
20 (patients)                          
& The FLIP device was inserted into the esophagus lumen, passing through the esophagus. The CSA at 16 location and one distal pressure measurements were taken as a function of time at 10- to 40-mL volume. \\\hline

Carlson et al. 2015 \cite{Carlson2015}    
& LES
&22-mm diameter
16-cm long
& 10 (controls)
51 (patients)                          
& Same as procedure above with a longer bag, and experiment conducted at 50- and 60-mL volume. \\\hline

Gourcerol et al. 2015 \cite{gourcerol2015impaired}    
& PS
&25-mm diameter
12 cm long
& 21 (controls)
32 (patients)                          
& The catheter was inserted trans-nasally, and the bag was positioned such that it passes through the pylorus. The bag volume was set to 10, 20, 30, and 40 mL. Pressure and CSA were reported at stable condition (not over time) so peristaltic effects on their values are not captured. \\\hline

Malik et al. 2015 \cite{malik2015assessing}    
& PS
&EndoFLIP device
& 54 (patients)                          
& The catheter was inserted orally, and the bag was positioned such that it passes through the pylorus. The bag volume was set to 20, 30, 40, and 50 cc. The recorded P and CSA were taken from a single time instance. \\\hline

Kunwald et al. 2010 \cite{kunwald2010new}    
& SoO
&9-mm diameter
5-cm long
& 4 (patients)                          
& The probe was inserted into the SO and the bag was filled at a rate of 1 ml/min. During this process, the CSAs and pressures were recorded.\\\hline

Zifan et al. 2019 \cite{Zifan2019} 
& AS
&30-mm diameter
90-mL maximal volume
& 14 (F - controls)
14 (F - patients)      
& FLIP bag was placed in the AS. Three, 10 sec maximum voluntary squeezes were performed at each bag volume (50,70,90 mL). \\\hline

Gourcerol et al. 2016 \cite{gourcerol2016endoflip}      
& AS
&25-mm diameter
12-cm long
& 40 (F - controls)
34 (F - patients)                            
& FLIP was inserted in the rectum. The bag volume was set to 10, 20, 30, 40, or 50 mL and the participants were asked to squeeze the filled bag. \\\hline

Klarskov and Lose 2007  \cite{klarskov2007urethral} 
& US-F
&5-mm diameter
6-cm long
& 143 (subjects)    
& The catheter surrounded by an elastic, cylindrically shaped bag was placed inside the urethra, passing through the sphincter. The bag volume increased systematically by a controlled, stepwise, inflation and deflation of the bag with air. The P and CSA at each time increment was recorded.  The experiment was conducted in both relaxed and squeezed state. \\\hline

Aagaard et al. 2012 \cite{aagaard2012urethral} 
& US-M
&7.5-mm diameter
5-6-cm long
& 10     
& Similar experimental set up as described above. \\
        \bottomrule
    \end{tabular}
    \label{table:LitList}
\end{table}

From figure \ref{fig:litLoops}, we can identify two different loop types: 
\begin{enumerate}[\bfseries (i)]
  \item \textbf{Negative slope loop.} Pressure increases as CSA of the sphincter decreases, such as in figure \ref{fig:AS_lit}. A similar pattern is seen in figure \ref{fig:SO_lit}.
\item \textbf{Positive slope loop.} Pressure increases as CSA of the sphincter increases, such as in figure \ref{fig:LES_lit}.
\end{enumerate}

\noindent If all sphincters relax and contract, why do we see different loop patterns (positive and negative slopes)? We aim to answer this question. We wish to investigate the different pressure-CSA loop patterns and reveal the underlying reason which leads to the different loops at different sphincters. We hypothesize that the loop types are a result of different muscle activities at the sphincter and the location upstream to the sphincter. In particular, we hypothesize that the loop type is dictated by the presence (or absence) of two mechanical modes discussed next.

Table \ref{table:openingModes} lists and describes the opening and closing mechanisms of each sphincter as obtained from literature survey. There are two potential mechanisms to open sphincters:

\begin{enumerate}[\bfseries (i)]
  \item Neurogenic mediated relaxation of the sphincter itself
  \item Peristaltic contraction or some squeezing upstream of the sphincter that causes mechanical distention.
\end{enumerate}

\noindent A sphincter's opening function can be defined by either only mechanism (i) (Case 1), or a combination of (i) and (ii) acting together (Case 2). Case 1 refers to sphincters which open only because of tonic relaxation of the sphincter muscle, like the AS \cite{moore2018clinically,schuster1963internal}. This neurally controlled muscle activity can be either voluntary or involuntary. Case 2 refers to sphincters which open as a result of the two mechanisms. While the sphincter muscles relax, there is also peristalsis of some sort proximal to the sphincter which increases the pressure, causing mechanical distention, like in the LES \cite{Elisha2021Loop,Mittal1997,liebermann1998anatomic}. Note that for the second mechanism to be considered, the pressure in the lumen must play a role in the sphincter's opening. The anal canal for instance is a high-pressure zone which increases as the canal is getting filled. However, this increase in pressure triggers the internal sphincter to relax but is not mechanically involved in the opening process itself  \cite{meunier1977control,burleigh1983neural}. 

Note that different from other sphincters, the upper esophageal sphincter (UES) does not have a resting tone, but rather responds to a vast array of reflexive inputs. It does not open due to muscle relaxation nor mechanical distention, but by anterior traction on the cricoid cartilage, which forces the sphincter to open \cite{Kahrilas2022}. The UES opening is a result of pulling applied by external forces. With that said, the mechanical imprint is equivalent to relaxation of the tone. Thus, we classify the UES's opening by mechanism (i) (Case 1). We will later show that whether the sphincter opens due to neurogenic mediated relaxation or due to neurally controlled pulling of the sphincter wall, the pressure-CSA loop type is not affected.


\begin{table}[!htb]
\fontsize{8}{10}\selectfont
\centering
   \caption{Sphincters and their opening modes}
    \label{crouch}
    \begin{tabular}{  p{0.1\textwidth} p{0.65\textwidth}  p{0.15\textwidth}}
        \toprule
\textbf{Sphincter}      
& \textbf{Process Description}   
& \textbf{Mechanism} \\\midrule
UES 
& During swallowing, the UES opens through anterior traction on the cricoid cartilage by the strap muscles suspending the hyoid bone \cite{Kahrilas2022,Sivarao2000,hirano2015esophagus}. Once the UES is open, the bolus is transferred through the UES into the esophagus by tongue pulsion followed by peristaltic contraction \cite{Sivarao2000}.        
& (i) \\\hline
LES     
& After the bolus enters the esophagus through the UES, a peristaltic contraction pushes it towards the LES \cite{Hershcovici2011}. The peristaltic activity, together with active tone relaxation of the LES muscle open the LES and allow the emptying of the esophagus content into the stomach \cite{Elisha2021Loop,Mittal1997,liebermann1998anatomic}.                         
& (i) + (ii) \\\hline
PS    
& During the digestive state, antral contractions push the food against the constricted pylorus which only allows the passage of small particles \cite{Hellstrom2006}. The remainder of the chyme is propelled backwards and is subject to additional antral contractions. Only during the inter-digestive period, the migrating motor complex is activated to clear indigestible material from the stomach \cite{Deloose2012} Intragastric pressure is developed by a peristaltic wave passing through the antrum, causing the pylorus to open sufficiently wide to allow that material to exit \cite{hunt1983mechanisms, shafik2006mechanism}.
& (i)+(ii) (during inter-digestive) \\\hline
SoO &
There is no musculature in the bile duct or the pancreatic duct. Consequently, the pumping action is done by the sphincter itself. This can be triggered by the injection or secretion of cholecystokinin. Cholecystokinin triggers propagated contractions within the sphincter that milk bile or pancreatic secretions into the duodenum \cite{woods2005sphincter,Kuussayer1995}. Hence, the opening occurs by the sphincter itself relaxing with some help from pressurization.       
& (i)+(ii) (small effect from (ii)). \\\hline
IAS
&Once enough fecal material accumulates the rectum, it causes the rectum to distend, triggering the sphincter to tonically relax (nerval response) \cite{moore2018clinically,schuster1963internal,meunier1977control}. Peristalsis is not involved, and the opening does not depend on mechanical distention \cite{meunier1977control,burleigh1983neural}.
&(i) \\\hline
EAS
&When the IAS relaxes (involuntarily), it signals the need for the body to defecate \cite{moore2018clinically}. The EAS remains contracted, even after the IAS opens, unless a voluntary signal to defecate is in place \cite{moore2018clinically}. Once the signal to relax is received, the anorectal angle decreases, allowing the EAS to open \cite{moore2018clinically}. Following the opening, peristaltic waves and rectum contraction forcing fecal material out of the anal canal \cite{moore2018clinically}. This peristalsis does not play a mechanical role in the opening of the sphincter as it takes place after the sphincter has opened.
&(i) \\\hline
US
&Since the internal and external sphincters are interdependent, we refer to their opening and closing function as one unit, the US.
The sympathetic nervous system is responsible of keeping the US contracted, while the parasympathetic nervous system is responsible for relaxing the sphincter once the bladder is sufficiently full \cite{jung2012clinical,de1976reflex}. The relaxation of the muscle causes the sphincter to open and allow flow into the urethra \cite{moore2018clinically,jung2012clinical,de1976reflex}. Thus, the opening is neurally controlled. Peristalsis contraction force urine towards the external urethral orifice. However, the peristaltic activity at the urethra occurs distal to the urethral sphincter and therefore is not mechanically involved in the opening process of the sphincter. 
& (i)\\
        \bottomrule
    \end{tabular}
    \label{table:openingModes}
\end{table}


\section{Methods} \label{Method}
In this section, we discuss the experimental procedure to collect clinical FLIP data at the LES. In addition, we detail the simulation approach taken to reproduce FLIP experiments at different sphincters. 


\subsection{Subjects} \label{Subjects}
The subject cohort included 24 healthy, asymptomatic adult volunteers. This number excludes subjects with previous diagnosis of esophageal disorders, previous diagnosis of autoimmune, or eating disorders, use of antacids or proton pump inhibitors, body mass index $>30\text{ kg/m}^2$, or a history of tobacco use or alcohol abuse, as described in \cite{Carlson2021J}. The data was collected at the Esophageal Center of Northwestern between November 2012 and October 2018. Informed consent was obtained for subject participation; who were paid for their participation. The study protocol was approved by the Northwestern University Institutional Review Board. This control cohort overlaps previous publications \cite{Carlson2021J,Carlson2021,AcharyaEsoWork2020,Carlson2016,carlson2019normal,triggs2020functional,carlson2021evaluating}.

\subsection{FLIP Study Protocol} \label{StudyProtocol}

As described in \cite{Carlson2021J,Carlson2021,AcharyaEsoWork2020,Carlson2016,carlson2019normal}, the study was performed during sedated endoscopy using a 16-cm FLIP (EndoFLIP\textsuperscript{\tiny\textregistered} EF-322N; Medtronic, Inc, Shoreview, MN), presented in figure \ref{fig:FLIP_bag}. The FLIP was inserted transorally, placed in the esophagus lumen, bypassing the LES.  The FLIP bag was then filled with saline at 10-mL increments, and the CSA at 16 locations and one distal pressure measurements were recorded as the wall responded to the filling. The bag volume began at 40-mL and increased until reaching 60-mL. Each volume was maintained for 30-60 sec. More information on the experimental procedure is available in \cite{Carlson2021J,Carlson2021,AcharyaEsoWork2020,Carlson2016,carlson2019normal}.


\subsection{Data Analysis} \label{DataAnalysis}

The FLIP data was exported using MATLAB $\tt{readtable}$. A total of 219 contraction cycles were identified by tracking distal pressure and CSA at the LES, as described in \cite{Elisha2021Loop} and \cite{AcharyaEsoWork2020}. Due to dry catheter artifact (DCA), we omitted 87 data points. Consequently, the study focused on 132 contraction cycles. DCA is when the peristaltic contraction causes occlusion of the FLIP and therefore disrupts the electrical current which is used for the impedance planimetry technology. The pressure at different locations along the FLIP bag can be calculated using the provided distal pressure, as proposed by Halder et al. \cite{Halder_2021}. Hence, one can plot the pressure-CSA loops at different locations on the esophagus.

\subsection{Simulation of FLIP Experiment} \label{Simulation}

Given our understanding of the sphincters function from literature (table \ref{table:openingModes}), we can reproduce their opening and closing mechanism using simulations. Simulations allow us to dictate the activation function, and by that, imitate the muscle activity throughout a contraction cycle. Doing so, we can identify which muscle activation configurations result in each loop type.

In our previous studies, we developed a 1D model of a flow inside an elastic tube closed on both ends in order to imitate a flow inside a FLIP device \cite{Acharya_2021,Elisha2021,Elisha2021Loop}. Our goal was to study the relation between tube properties, fluid properties, muscle activation pattern, and their effect on pressure and CSA of the tube. In this section, we only include minimal details of the model and the simulation procedure. Extended information can be found in \cite{Acharya_2021,Elisha2021,Elisha2021Loop}.

\subsubsection{Governing Equations in 1D} \label{Governing_Equations}

The 1D mass and momentum conservation equations were derived in \cite{Ottesen2003} and take the form  
\begin{equation} \label{eq:continuity}
    \frac{\partial A}{\partial t}+\frac{\partial\left(Au\right)}{\partial x} = 0,
\end{equation}
and
\begin{equation} \label{eq:momentum}
    \frac{\partial u}{\partial t} + u\frac{\partial u}{\partial x} = 
    -\frac{1}{\rho}\frac{\partial P}{\partial x}-\frac{8\pi\mu u}{\rho A},
\end{equation}

\noindent  respectively. Here, $A(x,t)$, $u(x,t)$, $P(x,t)$,  $\rho$ and $\mu$ are the tube CSA, fluid velocity (averaged at each cross-sectional area), pressure inside the tube, fluid density, and fluid viscosity, respectively. In addition, we use a constitutive equation for pressure, which relates pressure and CSA, such that

\begin{equation}
{{P}}={K_{\scriptscriptstyle e}}\left(\frac{A(x,t)}{A_{\scriptscriptstyle o}\theta(x,t)}-1\right) + P_o,
\label{eqn:tube_law_theta}
\end{equation}

\noindent which was derived by \cite{Whittaker2010} and validated experimentally by \cite{Kwiatek2011}. In the equation above, ${P}_o$ is the outside pressure, $K_e$ is tube stiffness, and $A_o$ is the undeformed reference area (CSA of the tube when ${\Delta{P}}=P-P_o=0$) \cite{Acharya_2021,Elisha2021}. The term $\theta(x,t)$ is called the activation function, implemented to the model in order to mimic muscle contraction and relaxation by changing the reference CSA of the tube wall \cite{Acharya_2021,Bringley2008,Manopoulos2006}. When $\theta=1$, there is no contraction nor relaxation, and the system is fully at rest. When $\theta<1$, contraction is implemented. The smaller the value of $\theta$, the smaller the tube reference area, and the greater the contraction intensity \cite{Acharya_2021}. When $\theta>1$, relaxation is implemented \cite{Elisha2021}.

Recall that in the case of the UES, no muscle relaxation is present, but rather external traction which pulls the sphincter's wall to open. To simulate this scenario, we vary the value of $P_o$ over time, such that $ P_o= P_o(x,t)$ while keeping the reference CSA constant. That is, setting $\theta$ as constant over time (only a function of $x$). The relation in equation (\ref{eqn:tube_law_theta}) in this case is written as



\begin{equation}
{{P}}={K_{\scriptscriptstyle e}}\left(\frac{A(x,t)}{A_{\scriptscriptstyle o}\theta}-1\right) + P_o(x,t).
\label{eqn:tube_law_delta}
\end{equation}

\subsubsection{Peristaltic Wave Input and Active Relaxation} \label{Peristaltic_Wave}

In order to simulate contraction cycles during a FLIP experiment at the different sphincters, we wish to dictate a different muscle pattern for each case, based on the opening and closing modes discussed in section \ref{LitAnalysis}. To do so, we change the activation function $\theta$ in equation \ref{eqn:tube_law_theta}. Other than the muscle activity, the simulation set up is the same for all cases. The different activation functions are discussed next. Mathematical details on the functions themselves are available in \cite{Acharya_2021,Elisha2021Loop}.

\begin{figure*}

    \centering
    \begin{subfigure}[b]{0.24\textwidth}
        \centering
        \includegraphics[trim=30 130 60 175,clip,width=\textwidth]{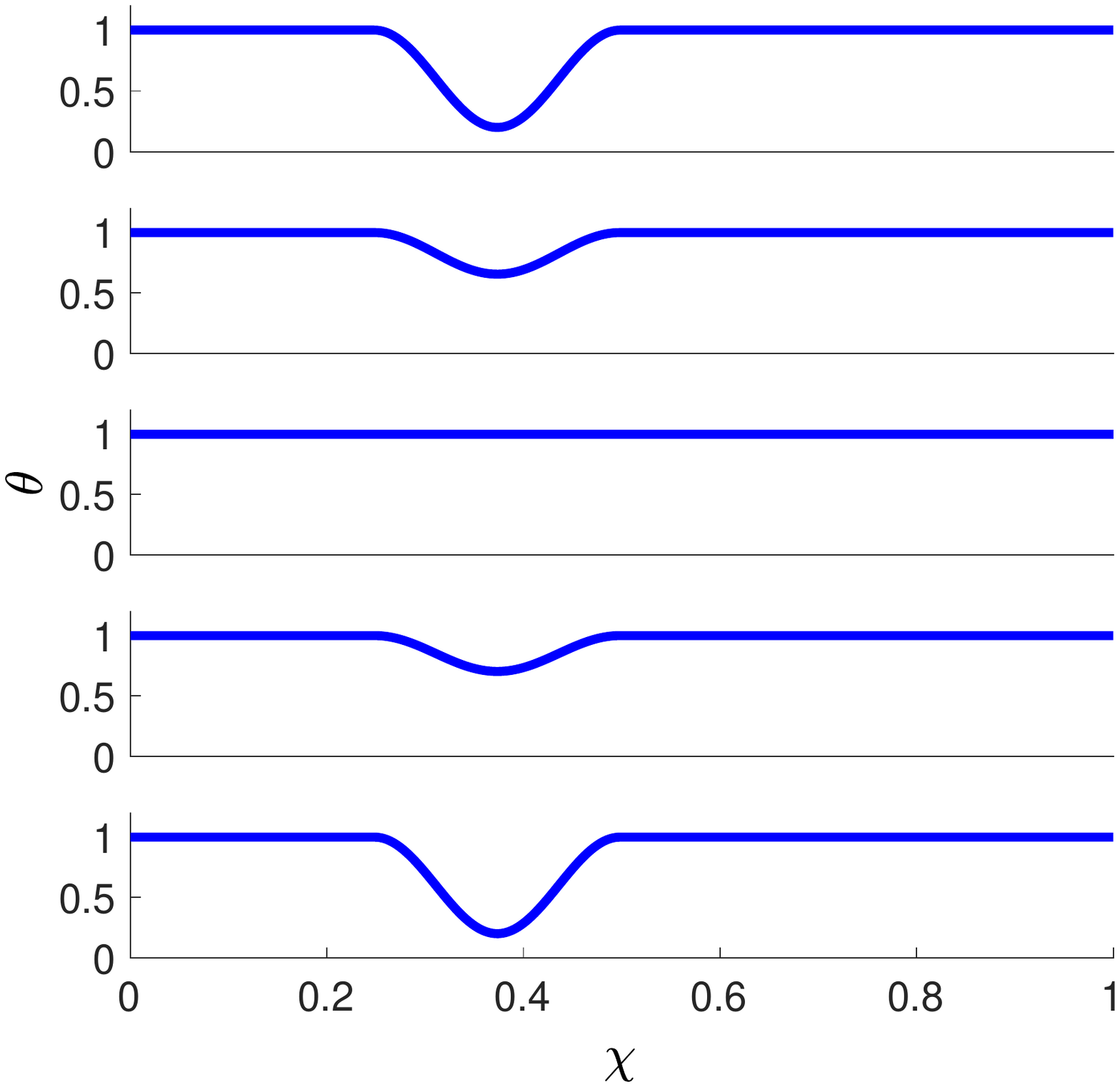}
        \caption{Tone relaxation}
        \label{fig:activationSphincter}
    \end{subfigure}
    \ 
    \begin{subfigure}[b]{0.24\textwidth}  
        \centering 
        \includegraphics[trim=30 130 60 175,clip,width=\textwidth]{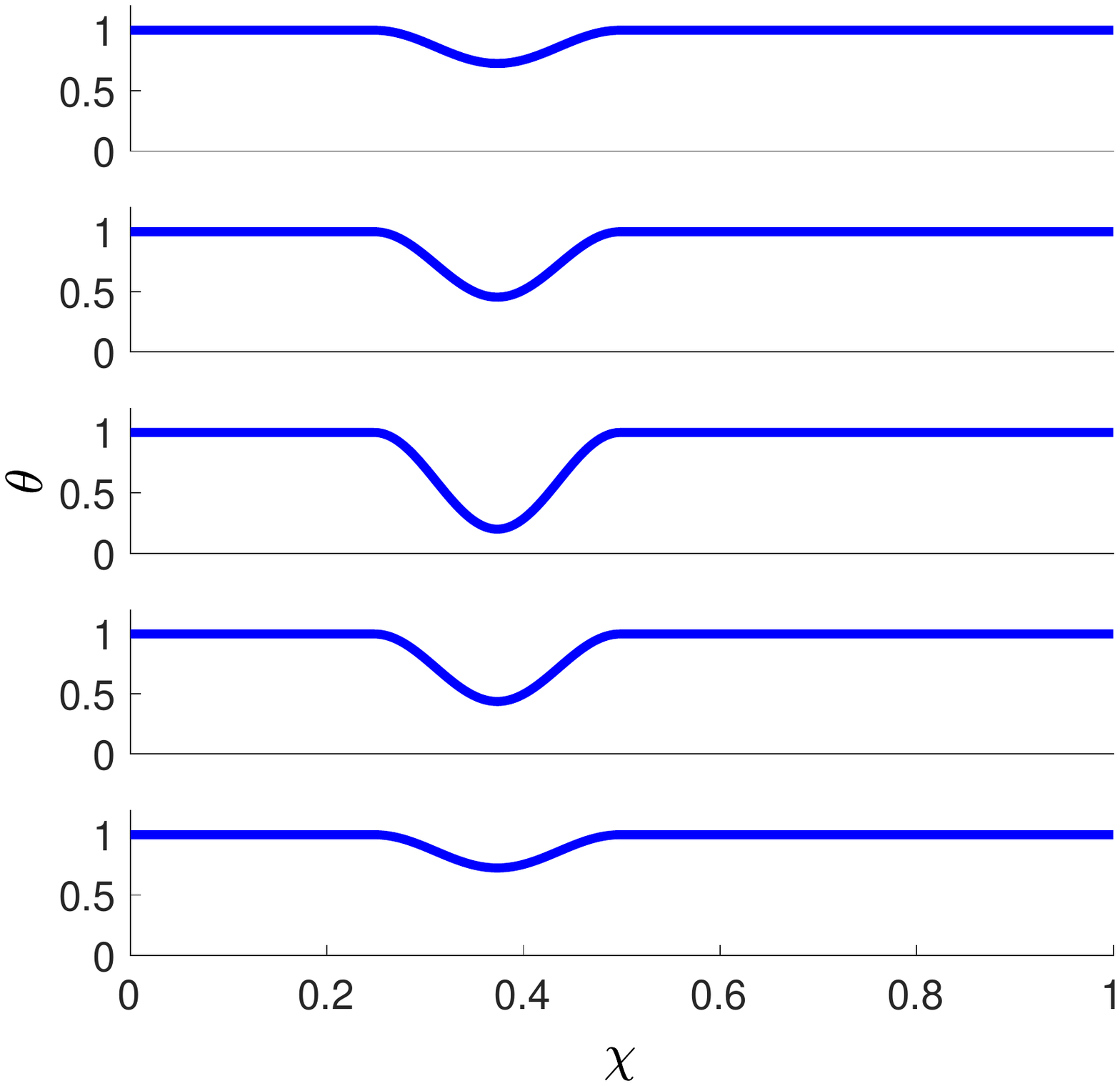}
        \caption{Tone squeezing}
        \label{fig:activationSqueez}
    \end{subfigure}
    \ 
    \begin{subfigure}[b]{0.24\textwidth}   
        \centering 
                \includegraphics[trim=30 130 60 175,clip,width=\textwidth]{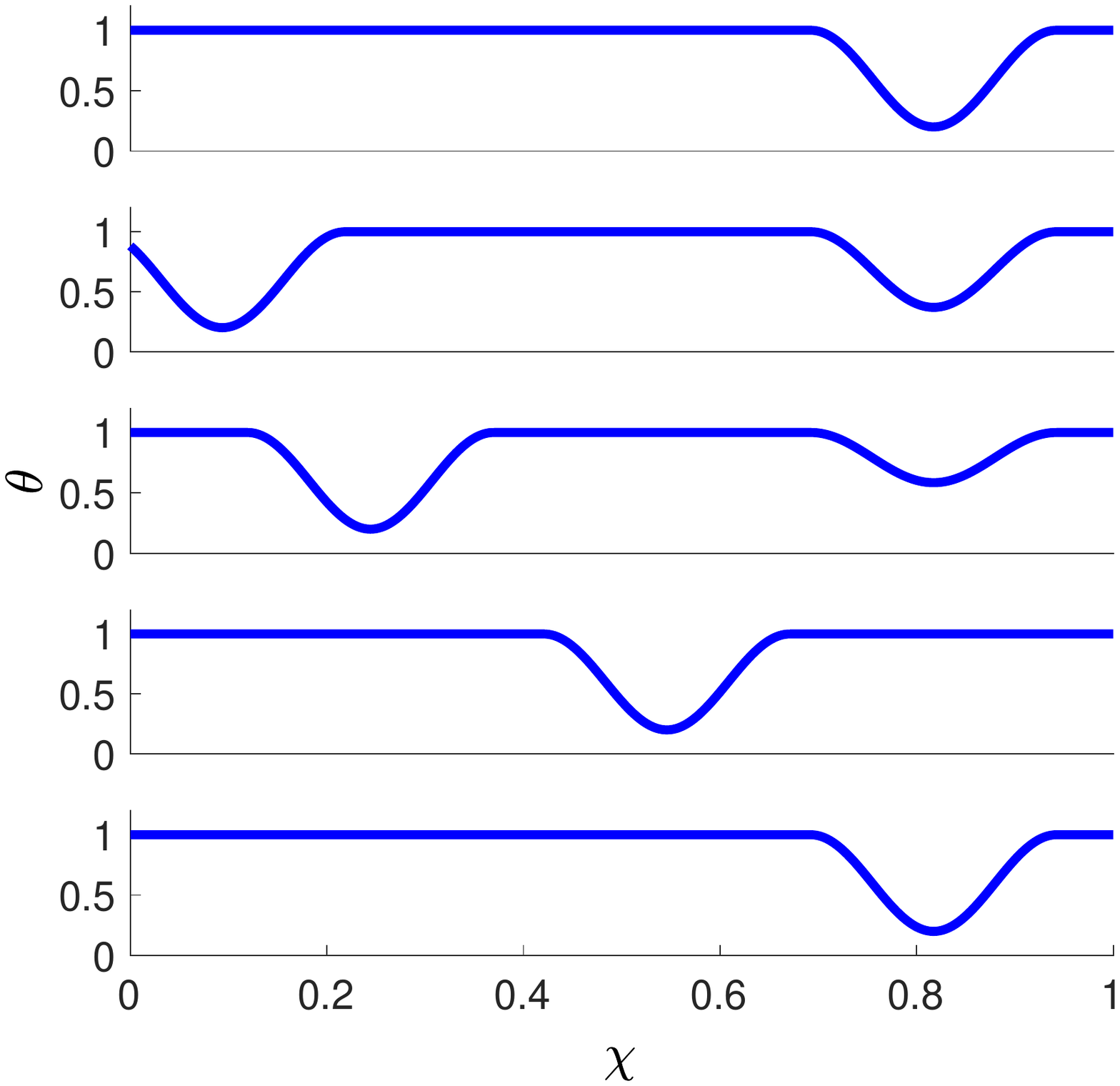}
        \caption{Tone and peristalsis}
        \label{fig:activationEGJ}
    \end{subfigure}
    \ 
    \begin{subfigure}[b]{0.24\textwidth}   
        \centering 
        \includegraphics[trim=30 130 60 175,clip,width=\textwidth]{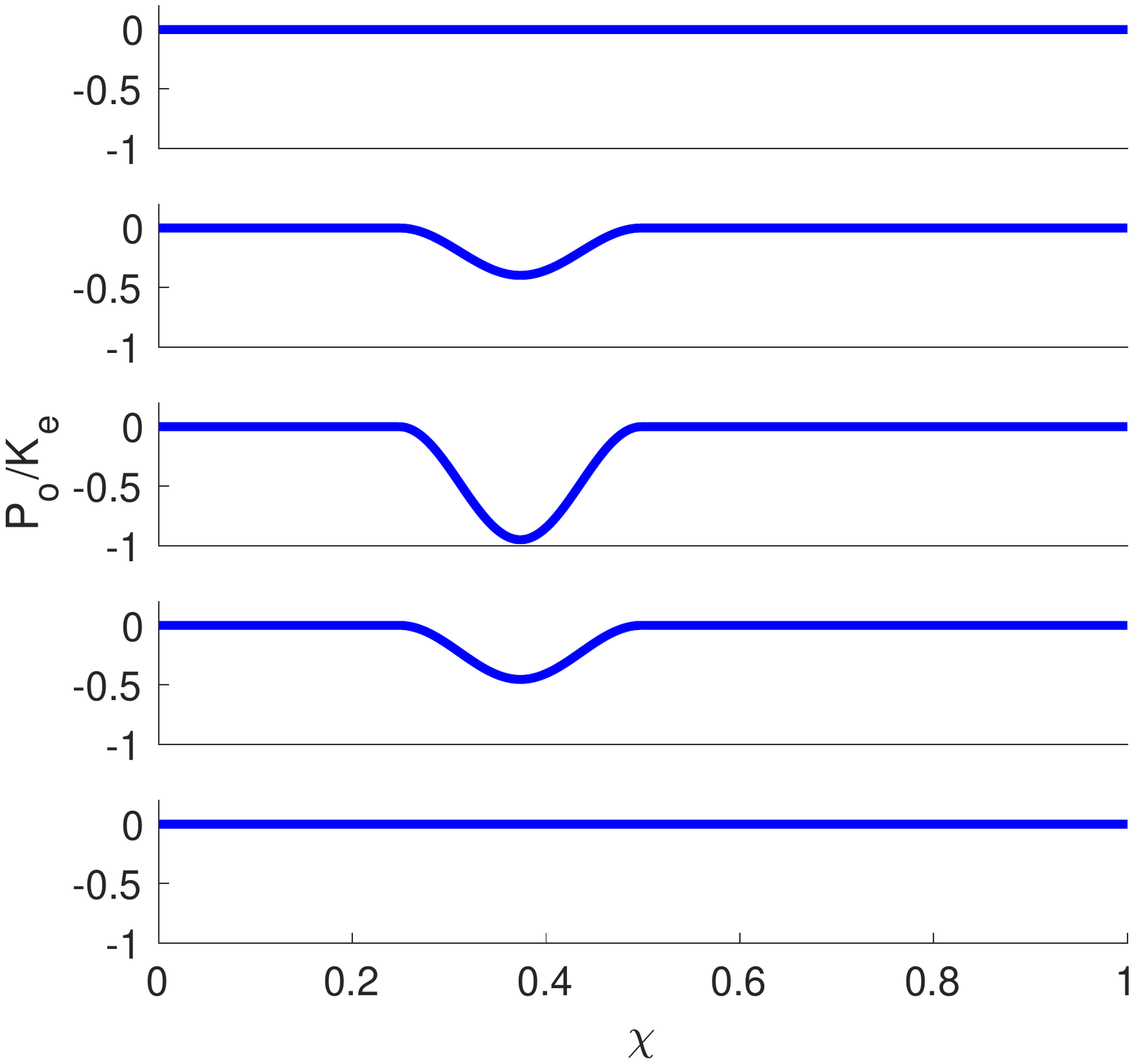}
        \caption{Outside pressure}
        \label{fig:PoFunct}
    \end{subfigure}
    \caption{Illustrations of the four activation functions used for the simulations in this study. Each subfigure presents five consecutive instances. Figures (a),(b), and (c) present different $\theta(x,t)$ functions while figure (d) presents $P_o(x,t)/K_e$.}
    \label{fig:activationFunctions}
\end{figure*}

\textbf{(a) Tone relaxation.}
This activation function simulates a contracted sphincter which relaxes and then contracts back as seen in the regular function of the AS of the US. The active relaxation of the sphincter is the only mode mechanically involved in the opening of the sphincter. Figure \ref{fig:activationSphincter} presents this activation pattern at five consecutive time instances.

\textbf{(b) Tone squeezing.}
This activation function simulates a scenario similar to the experimental procedure conducted by Zifan et al. \cite{Zifan2019}, in which the partially relaxed sphincter contracts and then relaxes back again. This function imitates active squeezing of the FLIP bag, capturing a scenario opposite of the regular function of the AS. Figure \ref{fig:activationSqueez} presents this activation pattern at five consecutive instances.

\textbf{(c) Tone relaxation and traveling peristalsis.}
This activation function involves both peristaltic contraction and a tonically contracted sphincter, capturing the two mechanical modes, and is discussed in details in \cite{Elisha2021Loop}. As the traveling peristalsis enters the domain, causing a pressure rise in the tube, the contracted sphincter starts to relax. The contraction of the tone at the end of the contractile cycle is obtained by the traveling peristalsis stopping at the sphincter location \cite{Elisha2021Loop}. This pattern is seen in the regular function of the LES. Figure \ref{fig:activationEGJ} presents this activation pattern at five consecutive instances.

As discussed in section \ref{Governing_Equations}, we also want to introduce a simulation where the opening and closing of the sphincter is dictated by external traction, which can be achieved by varying $P_o$ over time. Figure \ref{fig:PoFunct} presents $P_o(x,t)$ at five consecutive time instances. Note that $P_o$ in this plot is non-dimensionalized by $K_e$, as discussed in the next section. In these type of simulations, $\theta=\theta(x)$, as the reference CSA does not change with time. The function $\theta(x)$ takes the form of $\theta$ at the first time instance presented in figure \ref{fig:activationSphincter}. In the simulations described in \textbf{(a)}-\textbf{(c)} above, $\theta=\theta(x,t)$ and $P_o=0$.

\subsubsection{Non-dimensionalizing Dynamic Equations} \label{nondimParametersGov}

We use the following scales to non-dimetionalize our equations:

\begin{equation}
A=\alpha A_{o}, \qquad t=\tau t_R, \qquad u=U\frac{L}{t_R}, \qquad P=pK_e, \qquad \text{and} \qquad x=\chi L,
\end{equation}

\noindent where $\alpha$, $\tau$, $U$, $p$, and $\chi$ are non-dimensional variables of area, time, velocity, pressure, and position, respectively \cite{Acharya_2021,Elisha2021}. The terms $L$ and $t_R$ are dimensional constants for the the tube length and relaxation time, respectively. 

Hence, the non-dimetionalized form of the mass conservation, momentum conservation, and tube law equations are
\begin{equation} \label{eq:continuity_nondim}
    \frac{\partial\alpha}{\partial\tau}+\frac{\partial\left(\alpha U\right)}{\partial\chi} = \epsilon\left(\frac{\alpha}{\theta}\right)_{xx}, 
\end{equation}

\begin{equation} \label{eq:momentum_nondim}
    \frac{\partial U}{\partial\tau} + U\frac{\partial U}{\partial\chi} + 
    \psi\frac{\partial p}{\partial\chi}
    + \beta\frac{U}{\alpha} = 0, \qquad \text{and}
\end{equation}

\begin{equation} \label{eq:tube_law_nondim}
    \textcolor{REDCOLOR2}{p=\left( \frac{\alpha}{\theta}-1 \right)+p_o- \eta\frac{\partial\left(\alpha U \right)}{\partial\chi}},
\end{equation}

\noindent respectively \cite{Acharya_2021,Elisha2021}. In the equations above, $\psi=K_e{t^2_R}/(\rho L^2)$ is the non-dimensional stiffness parameter, $\beta = 8\pi\mu t_R/(\rho A_o)$ is the non-dimensional viscosity parameter, and $p_o=P_o/K_e$ is the outside pressure. Notice that the equations above include two terms which were not included in the dimensional form of the equations. First, the term $\epsilon\left(\frac{\alpha}{\theta}\right)_{xx}$ is a smoothing term added to the right-hand side of the continuity equation in order to obtain faster convergence and reduce computational time \cite{LeVeque1990}. The expression $\eta\frac{\partial\left(\alpha U \right)}{\partial\chi}$ is added to the pressure equation in order to regularize the system and therefore help stabilize the numerical solution, where $\eta=(YA_o)/(t_RK_e)$, and Y is a damping coefficient \cite{Wang2014}. The effect of implementing these term into the model is minimal as discussed in \cite{Acharya_2021}.

\subsubsection{Boundary and Initial Conditions} \label{Boundary_Conditions}

The FLIP bag is closed on both ends and the volume inside it remains constant, hence,
\begin{equation} \label{eq:velBC}
    U\left(\chi=0,\tau\right)=0\qquad\text{and}\qquad U\left(\chi=1,\tau\right)=0.
\end{equation}
Plugging this into equation (\ref{eq:momentum_nondim}), and taking the partial derivative of equation (\ref{eq:tube_law_nondim}) yields a Neumann boundary condition for $\alpha$, such that 
\begin{equation} \label{eq:areaBC}
    \left.\frac{\partial}{\partial \chi} \left(\frac{\alpha}{\theta}+p_o\right)\right|_{\chi=0,\tau}=0\qquad\mathrm{and}\qquad
    \left.\frac{\partial}{\partial \chi} \left(\frac{\alpha}{\theta}+p_o\right)\right|_{\chi=1,\tau}=0
\end{equation}
\cite{Acharya_2021,Elisha2021,Elisha2021Loop}. 

At $\tau=0$, the fluid inside the tube is at rest, so
\begin{equation} \label{eq:velIC}
    U\left(\chi,\tau=0\right)=0.
\end{equation}
\noindent Moreover, the initial CSA of the tube is defined as
\begin{equation} \label{eq:areaIC}
    \alpha\left(\chi,\tau=0\right)=S_{\text{IC}}\theta(\chi,\tau=0),
\end{equation}
where $S_{\text{IC}}$ is a constant area that depends on the volume of the bag and $\theta_{\text{IC}}=\theta(\chi,\tau=0)$. Lastly, $p_o=0$ for all $\chi$ and $\tau$ unless specified otherwise.

\subsubsection{Numerical Implementation} \label{Numerical_Implementation}

By plugging equation (\ref{eq:tube_law_nondim}) into equation (\ref{eq:momentum_nondim}), we obtain a system of two equations which, together with the boundary and initial conditions in equations (\ref{eq:velBC}), (\ref{eq:areaBC}), (\ref{eq:velIC}), and (\ref{eq:areaIC}), can be used to solve for $\alpha(\chi,\tau)$ and $U(\chi,\tau)$. The MATLAB $\tt{pdepe}$ function is used to acquire the numerical solution, as described in greater details by \cite{Acharya_2021}, who also validated the 1D model by comparing it to an equivalent 3D immersed boundary simulation. 

The simulations are differentiated by both the activation function ($\theta$ function described in section \ref{Peristaltic_Wave}) and a unique combination of $\psi$ and $\beta$, the physical parameters defining this problem. Varying these parameters is equivalent to examining the effect of tube wall stiffness, fluid density, and fluid viscosity on the tube deformation \cite{Acharya_2021}. Their values span between 100-10,000 to capture all possible scenarios. In the clinical esophagus FLIP experiment performed in this study $\beta$ is order 1 and  $\psi$ is order $10^3$. In this scenario, viscous effects are negligible. Thus, in simulations, large $\beta$ and small $\psi$ values are also considered.

\subsection{Work Decomposition} \label{Work_MathDetails}

The pressure-area plots presented in section \ref{LitAnalysis} show that during an opening and closing cycle, the opening and closing curves take different paths on the pressure-CSA axis, which creates a loop. This implies that there is some energy that is being gained or lost by the sphincters' wall. Hence, we wish to study the way in which energy is spent during a contraction cycle for the different scenarios. The conservation of work equation to conduct such analysis has been derived by \cite{Acharya_2021} and \cite{Elisha2021Loop}, and is briefly discussed here.

The conservation of work equation is obtained by integrating the momentum equation with respect to $\chi$ and $\tau$, such that 

\begin{equation}  \label{eq:work_balance_NonDim_no_split}
\begin{aligned}
{-\psi\int\limits _{\tau_1}^{\tau_2}\int\limits _{\chi_1}^{\chi_2}p\frac{\partial \alpha}{\partial \tau}\mathrm{d}\chi\mathrm{d}\tau} &={\int\limits _{\tau_1}^{\tau_2}\frac{\partial}{\partial \tau}\int\limits _{\chi_1}^{\chi_2}\left(\frac{1}{2}\alpha U^{2}\right)\mathrm{d}\chi\mathrm{d}\tau} +
{\beta\int\limits _{\tau_1}^{\tau_2}\int\limits _{\chi_1}^{\chi_2}U^{2}\mathrm{d}\chi\mathrm{d}\tau} \\
& +{\psi\int\limits _{\tau_1}^{\tau_2}\left(\alpha Up\right)\bigg|_{\chi_1}^{\chi_2}\mathrm{d}\tau} + 
{\int\limits _{\tau_1}^{\tau_2}\left(\frac{1}{2}\alpha U^3\right)\bigg|_{\chi_1}^{\chi_2}\mathrm{d}\tau}
 \end{aligned}
\end{equation}
\cite{Acharya_2021,Elisha2021Loop}. The time integration boundaries, $\tau_1$ and $\tau_2$, are the start and end of the contraction cycle. The spacial integration boundaries, $\chi_1$ and $\chi_2$ mark the start and end locations of the sphincter along the tube length, where $\text{w}_{\text{s}}=\chi_2-\chi_1$ is the width of the sphincter. The left hand side of the equation represents the work done by the tube wall on the fluid, and the right hand side represents the consumers of this work. The consumers, in the order in which they appear in the equation, are kinetic energy of the fluid, energy loss due to viscous dissipation in the sphincter region, work done by the fluid inside the sphincter region of the fluid outside through pressure acting on the cross-section at the two ends of the sphincter, and net momentum flux. 

The work done by the sphincter wall on the fluid is a result of both passive active work done by the sphincter wall on the fluid \cite{Acharya_2021,Elisha2021Loop}. The passive and active work decomposition can be applied to equation (\ref{eq:work_balance_NonDim_no_split}), such that

\begin{equation}  \label{eq:work_balance}
\begin{aligned}
{-\psi\int\limits _{\tau_1}^{\tau_2}\int\limits _{\chi_1}^{\chi_2}p_{\text{active}}\frac{\partial \alpha}{\partial \tau}\mathrm{d}\chi\mathrm{d}\tau} &={\int\limits _{\tau_1}^{\tau_2}\frac{\partial}{\partial \tau}\int\limits _{\chi_1}^{\chi_2}\left(\frac{1}{2}\alpha U^{2}\right)\mathrm{d}\chi\mathrm{d}\tau} +
{\beta\int\limits _{\tau_1}^{\tau_2}\int\limits _{\chi_1}^{\chi_2}U^{2}\mathrm{d}\chi\mathrm{d}\tau} \\
& +{\psi\int\limits _{\tau_1}^{\tau_2}\left(\alpha Up\right)\bigg|_{\chi_1}^{\chi_2}\mathrm{d}\tau} + 
{\int\limits _{\tau_1}^{\tau_2}\left(\frac{1}{2}\alpha U^3\right)\bigg|_{\chi_1}^{\chi_2}\mathrm{d}\tau} +
{\psi\int\limits _{\tau_1}^{\tau_2}\int\limits _{\chi_1}^{\chi_2}p_{\text{passive}}\frac{\partial \alpha}{\partial \tau}\mathrm{d}\chi\mathrm{d}\tau},
 \end{aligned}
\end{equation}

\noindent where the passive work is placed on the right hand side of the equation, representing the elastic energy stored in the tube wall.

Passive pressure captures the change in CSA while $\theta$ remains unchanged. The active pressure on the other hand, represents the change in the reference area as $\theta$ changes, when the CSA is fixed. The non-dimensional active and passive pressures are defined as

\begin{equation} \label{eq:ActivePassivePressure}
    p_{\text{active}}(\chi,\tau)=\alpha\left(\frac{1}{\theta(\chi,\tau)}-\frac{1}{\theta_{\text{IC}}(\chi)}\right)\qquad\text{and}\qquad p_{\text{passive}}(\chi,\tau)=\frac{\alpha(\chi,\tau)}{\theta_{\text{IC}}(\chi)}-1,
\end{equation}
respectively \cite{Elisha2021Loop}. In the simulation where $\theta=\theta(x)$ and $p_o=p_o(\chi,\tau)\neq0$, 

\begin{equation} \label{eq:ActivePassivePressureUES}
    p_{\text{active}}(\chi,\tau)=0\qquad\text{and}\qquad p_{\text{passive}}(\chi,\tau)=\frac{\alpha(\chi,\tau)}{\theta_{\text{IC}}(\chi)}-1+p_o(\chi,\tau).
\end{equation}

\section{Results} \label{Results}

\subsection{Lower Esophageal Sphincter Pressure-Area Loops} \label{LESLoops}

\begin{figure*}[!htb]
    \centering
    \begin{subfigure}[b]{0.480\textwidth}
        \centering
        {\includegraphics[trim=30 180 60 200,clip,width=\textwidth]{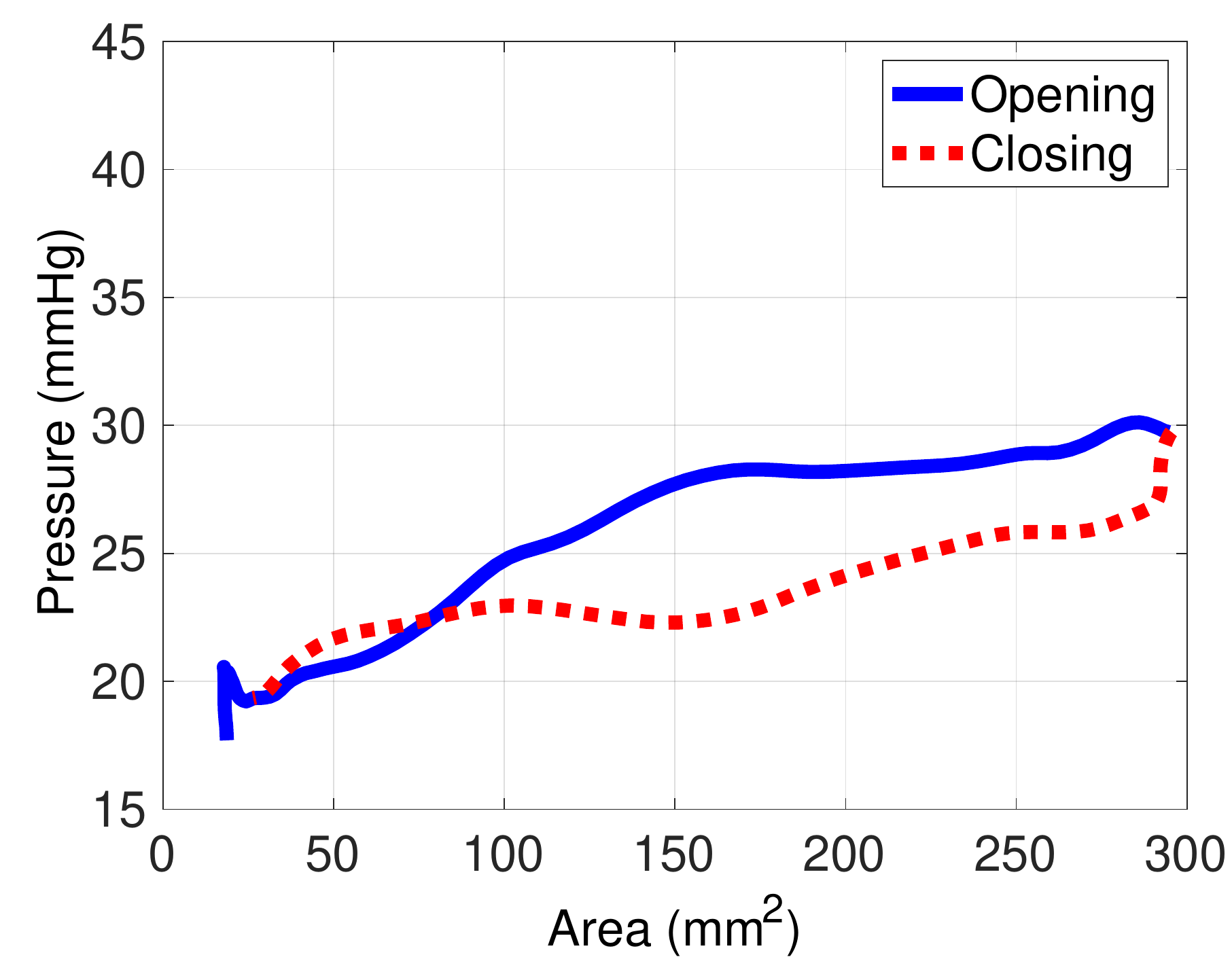}}
        \caption{Pressure dominant loop}
        \label{fig:PDL_clinical}
    \end{subfigure}
    \hfill
    \begin{subfigure}[b]{0.480\textwidth}  
        \centering 
        {\includegraphics[trim=30 180 60 200,clip,width=\textwidth]{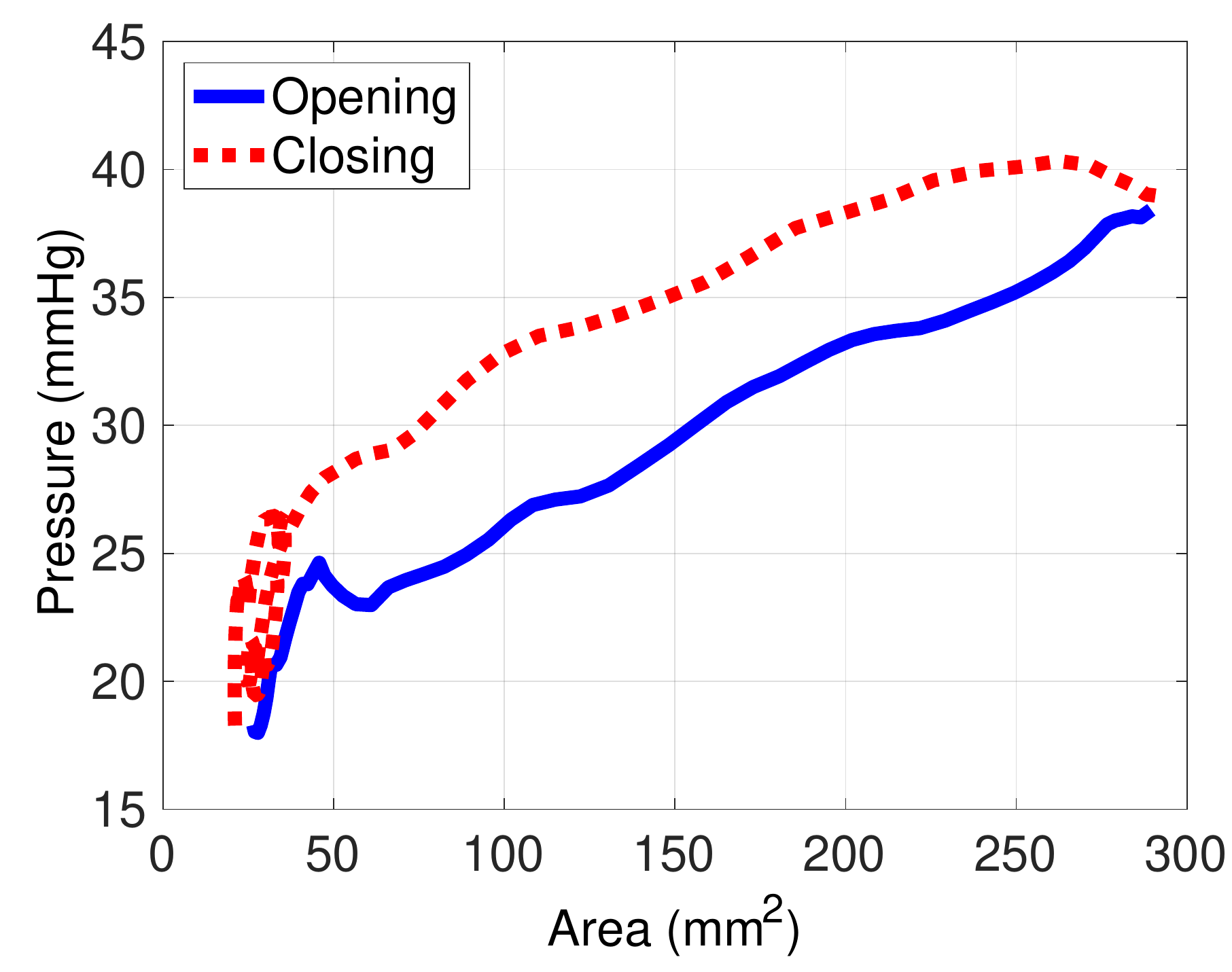}}
        \caption{Tone dominant loop}
        \label{fig:TDL_clinical}
    \end{subfigure}
    \caption{Two clinically observed pressure-area loops at the LES.} 
    \label{fig:clinical_Loops}
\end{figure*}

Figure \ref{fig:clinical_Loops} presents two pressure-CSA loops at the LES extracted from clinical FLIP data during two different contraction cycles. The opening and closing curves are identified on these plots. As the figure shows and mentioned in section \ref{LitAnalysis}, these loops have a positive slope, where pressure and CSA increase and decrease together. The reason behind this observation is explained in this section.

In addition to the pressure-CSA decreasing-increasing pattern, the resulting loops at the LES reveal two major pressure-CSA loop types in controls. The first loop type, presented in figure \ref{fig:PDL_clinical} is the pressure dominant loop (PDL), where the opening curve is above the closing curve. This loop appears in 34\% of the contraction cycles. The second loop type, presented in figure \ref{fig:TDL_clinical} is the tone dominant loop (TDL), where the closing curve is above the opening curve. This loop appears in 55\% of the contraction cycles. The names of these loops were introduced by Elisha et al. \cite{Elisha2021Loop}, and they capture the mechanical difference between the two loop types, which is discussed briefly in section \ref{PosSlope}. The remaining 11\% are contraction cycles which displayed the same pressure-CSA increasing-decreasing relation (positive slope), but a clearly defined loop is not obtained (the opening and closing curves cross one another).

\subsection{Negative Slope Pressure-Cross-Sectional Area Loop} \label{NegSlope}


Our simulation results show that when the only mechanism to open the sphincter is the neurogenic mediated relaxation (mechanism (i)), a negative slope loop (NSL) (figure \ref{fig:AS_lit}) emerges. Similarly, when the only mechanism to open the sphincter is an external force (also mechanism (i)), a negative slope loop (NSL) (figure \ref{fig:AS_lit}) is obtained. On the other hand, when the sphincter opens due to both neurogenic mediated relaxation of the tone and mechanical distention (mechanisms (i) and (ii)), a positive slope loop (PSL) (figure \ref{fig:LES_lit}) occurs. Hence, we show that the different pressure-CSA loop patterns originate from different neurally controlled muscle activity, and that the loop type is dictated by the presence or absence of the two mechanical modes proposed in section \ref{LitAnalysis}. 


In the rest of section \ref{Results}, we aim to explain the above statement by looking at different simulation results, separated into two categories, based on their resulting loop type; NSL (figure \ref{fig:AS_lit}), and PSL (figure \ref{fig:LES_lit}). We investigate the different activation functions, examine what causes these loop types, and inspect how work is distributed in the system. 


As discussed in section \ref{Peristaltic_Wave}, two different scenarios were modeled to imitate cases in which only the tone muscle constitute the opening of the sphincters. The activation functions in these simulations are presented in figures \ref{fig:activationSphincter} and \ref{fig:activationSqueez}. This section discusses two of these cases, one with the activation function in figure \ref{fig:activationSphincter} (sphincter contracted $\rightarrow$ relaxed $\rightarrow$ contracted) and one with the activation function in figure \ref{fig:activationSqueez} (sphincter partly relaxed $\rightarrow$ contracted $\rightarrow$ partly relaxed).

\begin{figure*}[!htb]
    \centering{{\includegraphics[trim=0 0 0 5 ,clip,width=0.9\textwidth]{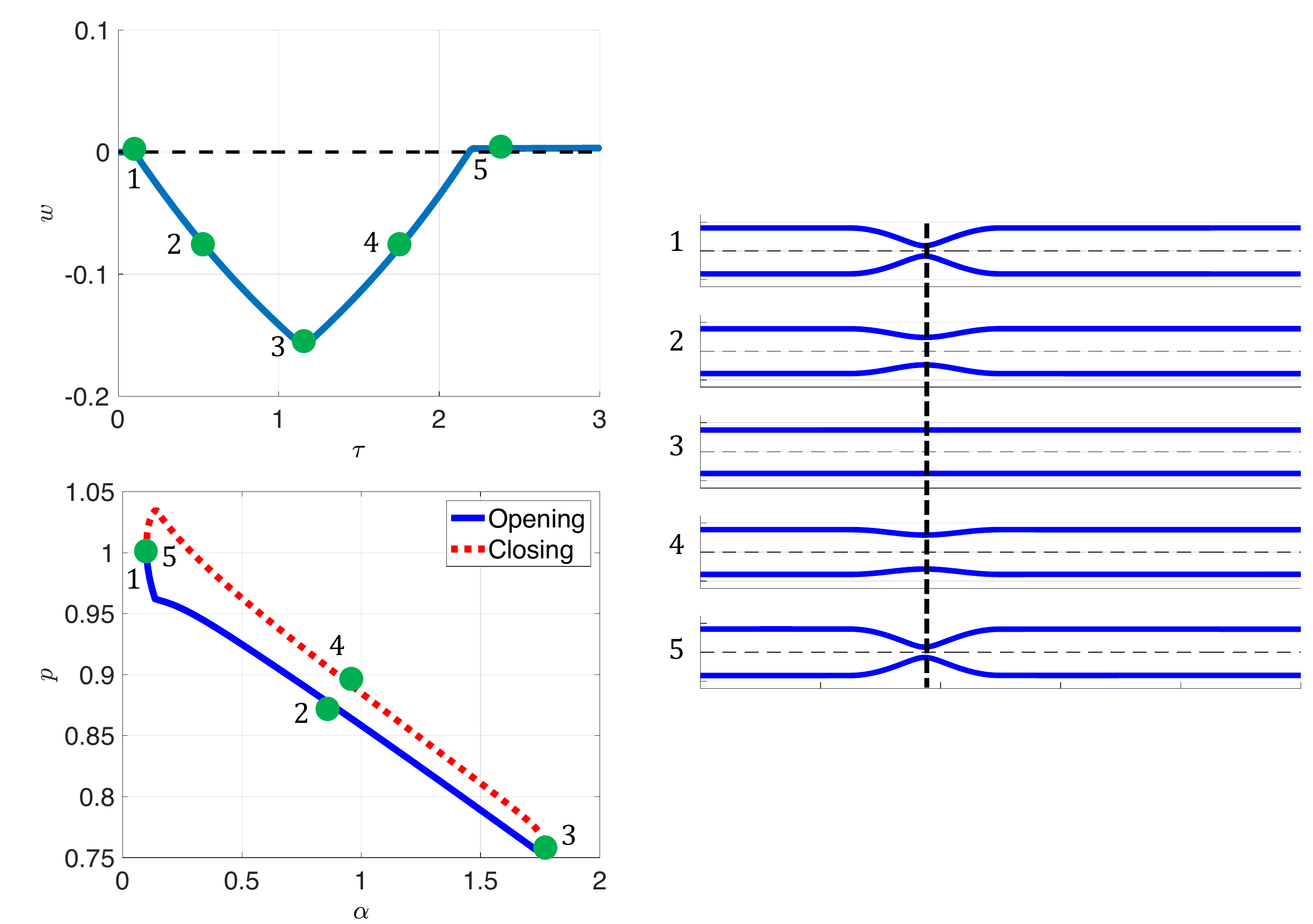}}}
    \caption{Simulation results of a single contraction cycle with muscle activity defined in figure \ref{fig:activationSphincter}. The graph at the top left shows the curve of the cumulative total work done by the sphincter wall on the fluid as a function of time. The plot at the bottom left presents the corresponding pressure-CSA loop as recorded at the sphincter location. The right plot displays the tube shape at five consecutive instants, ordered chronologically, corresponding to the five points highlighted on the two left plots.}
    \label{fig:simulationToneRelaxResults1}
\end{figure*}

Figure \ref{fig:simulationToneRelaxResults1} presents the results obtained by a simulation of a single contraction cycle with the activation function presented in figure \ref{fig:activationSphincter}. In this simulation, $\beta=100$, $\psi=100$, and the sphincter's width is $25\%$ of the tube length ($\text{w}_{\text{s}}=0.25$). The graph at the top left displays the total work done by the sphincter wall on the fluid (left hand side of equation \ref{eq:work_balance_NonDim_no_split}) normalized by $\psi$ ($w=\text{Work}/\psi$), as a function of time. The spatial integration is over the sphincter region. Note that the curve portrays the cumulative work done by the wall on the fluid up to a given time instant. The bottom left figure presents the corresponding pressure-CSA loop at the sphincter location. As the figure shows, a NSL is obtained. The five points marked on the loop and the work curve represent five time instants in the contraction cycle. The tube shapes at these five instants are plotted on the right. By doing so, we can relate the total work done by the sphincter wall on the fluid to the pressure-CSA loop, which helps us to better understand the pressure-CSA relation and consequently the loop type.

\begin{description}
  \item[\boldmath${\tau\leq\tau_1}$.]  At the beginning of the contraction cycle, the sphincter is contracted. At this time instance, the wall is still and the pressure is uniform, thus, the wall does no work on the fluid. As the work curve in figure \ref{fig:simulationToneRelaxResults1} shows, at this instance, the total work done by the wall on the fluid is equal to zero.
  
  \item[\boldmath${\tau_1<\tau<\tau_3}$.] Between $\tau_1$ and $\tau_3$, the sphincter tone starts to actively relax, causing the CSA at the sphincter to increase. This increase is seen in the loop in figure \ref{fig:simulationToneRelaxResults1} alongside pressure decrease. Since the sphincter opens, it does negative work on the fluid, as seen by the decrease in the total work curve in figure \ref{fig:simulationToneRelaxResults1} between $\tau_1$ and $\tau_3$.
    
  \item[\boldmath${\tau=\tau_3}$.] Once ${\tau=\tau_3}$, the sphincter reaches its full opening, as the loop in figure \ref{fig:simulationToneRelaxResults1} displays. Additionally, this instance corresponds to the minimum point on the work plot in the figure.
  
  \item[\boldmath${\tau_3<\tau<\tau_5}$.] Between instances $\tau_3$ and $\tau_5$, the sphincter starts to contract back to its original tone, causing the CSA at the sphincter to decrease. This decrease is seen in the loop in figure \ref{fig:simulationToneRelaxResults1} between points 3 and 5, alongside pressure increase. Since the contraction causes the sphincter to close, the sphincter squeezes the fluid, applying positive work on it. This process can be observed on the work plot in figure \ref{fig:simulationToneRelaxResults1} which shows that between points 3 and 5, the work done by the sphincter wall on the fluid increases.
  
   \item[\boldmath${\tau\geq\tau_5}$.] Lastly, the sphincter stops contracting once it reaches its original contraction strength. The final work value is equal to the net work done by the sphincter wall of the fluid during the entire cycle. 
   \end{description}

Notice that the net total work done by the sphincter wall on the fluid is positive. This indicates that throughout this opening and closing cycle, the sphincter wall applies more work on the fluid than the fluid (through pressure) applies on the wall. As elaborated upon by \cite{Elisha2021}, this loop type is called tone dominant loop (TDL), where the closing curve is above the opening curve on the pressure-CSA plot. Physically, it means that the sphincter wall needs to apply more work to contract back than the work needed to open the sphincter. This makes sense as, during closing, the wall needs to overcome fluid resistance, as elaborated upon later in this section. Note that this opening and closing pattern (closing curve above opening curve) appears in all simulations results with activation functions capturing only tone activity (figures \ref{fig:activationSphincter} and \ref{fig:activationSqueez}).



\begin{figure*}[!htb]
    \centering{{\includegraphics[trim=0 0 0 0 ,clip,width=0.9\textwidth]{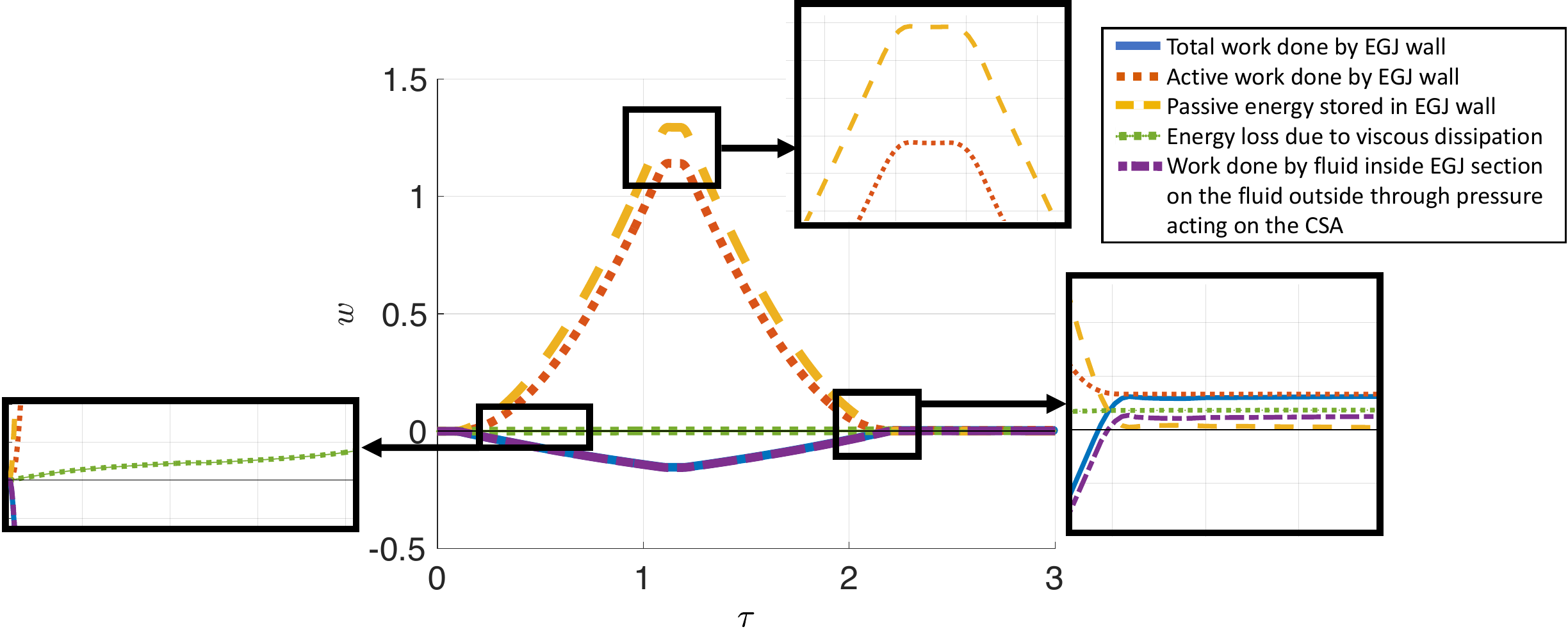}}}
    \caption{Plot of the work components from equation (\ref{eq:work_balance}) evaluated at the sphincter region for a simulation with muscle activity defined in figure \ref{fig:activationSphincter}. Each curve depicts the cumulative work done in the corresponding mode up to a given time instant. The total work done by the sphincter wall on the fluid (LHS of equation (\ref{eq:work_balance_NonDim_no_split}) is also plotted. }
    \label{fig:simulationToneRelaxResults}
\end{figure*}

Before proceeding our discussion on how the above analysis and segmentation relates to the different loop types identified in section \ref{LitAnalysis}, we will consider the detailed work balance in the sphincter region. Figure \ref{fig:simulationToneRelaxResults} presents the work components in equation (\ref{eq:work_balance}) and the total work term on the left hand side of equation (\ref{eq:work_balance_NonDim_no_split}), normalized by $\psi$ ($w=\text{Work}/\psi$), as a function of time. As the plot shows, at $\tau = 0$, before the contracted tone starts to relax, all the work components are equal to zero. However, once relaxation begins, all curves start decreasing or increasing, each having a unique pattern. Note that the rate of change of kinetic energy and the kinetic energy flux terms (first and fourth terms on the right hand side of equation (\ref{eq:work_balance}), respectively) are not plotted in the figure since $\psi>>1$ and $\beta>>1$ and consequently these contributions are small. In addition, recall that the spatial integration is over the sphincter region, and that each curve portrays the cumulative work done in the corresponding mode up to a given time instant.


The active work (the left hand side of equation (\ref{eq:work_balance})) is defined by the change in $\theta$. Hence, during relaxation, $\theta$ increases and the active work done by the sphincter wall on the fluid increases, as seen in figure \ref{fig:simulationToneRelaxResults}. As explained in \cite{Elisha2021Loop}, while the sphincter relaxes, the active pressure, as defined in equation (\ref{eq:ActivePassivePressure}) is negative, which implies that the fluid is sucking the sphincter's wall down. The direction of the force exerted by the wall on the fluid is same to the direction of the wall motion, so the active work is positive. The increase stops when no change in active work occurs ($\tau\approx1.20$), which corresponds to the instance in which the tone is fully relaxed ($\theta=1$). Once the sphincter starts to contract, $\theta$ decreases and the force exerted by the wall on the fluid is opposite to the direction of wall motion, so the active work is negative. This decrease stops when relaxation ends, where $\theta=\theta_{\text{IC}}$.

Passive work (last term on the right hand side of equation (\ref{eq:work_balance})) captures the change in CSA. Since only tone relaxation controls the change in CSA in this simulation, increase, decrease and maximum relaxation occurs simultaneously to increase, decrease and maximum CSA (full opening), respectively. Hence, the passive work curve must be very close to the active work curve, both in value and trend. However, in contrast to the active work, the net passive energy stored in the sphincter wall must equal to zero since it is reversible.

The work done by the fluid inside the sphincter section on the fluid outside of the sphincter through pressure acting on the CSA at the two ends of the sphincter region (third term on the right hand side of equation (\ref{eq:work_balance})) decreases as the sphincter relaxes. This is because as the sphincter relaxes and opens, fluid from outside of the sphincter region flows into the sphincter section, which does positive work on the fluid inside. As the sphincter contracts back, an opposite affects occurs, where the contraction of the wall forces the fluid occupying the sphincter region to flow out of the sphincter region, applying positive work on the outside fluid. Notice that the value of the net work of this component is positive, meaning that fluid inside the sphincter region does more work on the fluid outside than the opposite. 

The energy loss due to viscous dissipation (second term on the right hand side of equation (\ref{eq:work_balance})) maintains a steady increase throughout the contraction cycle. Since viscous dissipation cannot be recovered, the net energy that is being dissipated is positive. As the sphincter opens, viscous losses increase as a result of fluid flowing into the sphincter region. As the sphincter closes, viscous losses increase as a result of fluid flowing out of the sphincter region. 

Understanding the trends of the viscous losses and the work done on the fluid within the sphincter region by the pressure imposed on the cross-sections at the two ends of the sphincter region (negative of the third term on the right hand side of equation (\ref{eq:work_balance})) is particularly useful. They explain why the net total work done by the sphincter wall on the fluid is positive. During opening, fluid outside of the sphincter region flows into the sphincter region by pressure gradient. On the other hand, during closing, the wall itself applies the work to push the fluid occupying the relaxed sphincter segment away through the two ends of the sphincter region, while overcoming fluid resistance.

The above discussion and the time decomposition into the five stages presented in figure \ref{fig:simulationToneRelaxResults1} explain the shape of the loop. In this simulation, tone activity is the only component that controls the change of CSA of the sphincter, with relaxation and contraction corresponding to opening and closing of the sphincter, respectively. Pressure change is directly related to the change in the shape of the elastic tube. Thus, with no additional pressurization source in the form of contraction elsewhere along the tube length, tone activity is the only stimulus that results in pressure change in the tube. By elementary physics, if no other pressurization source is applied as the sphincter relaxes and the CSA increases, then pressure must drop. As the sphincter contracts, CSA decreases while pressure increases back up.

\begin{figure*}[!htb]
    \centering{{\includegraphics[trim=0 0 0 0 ,clip,width=0.9\textwidth]{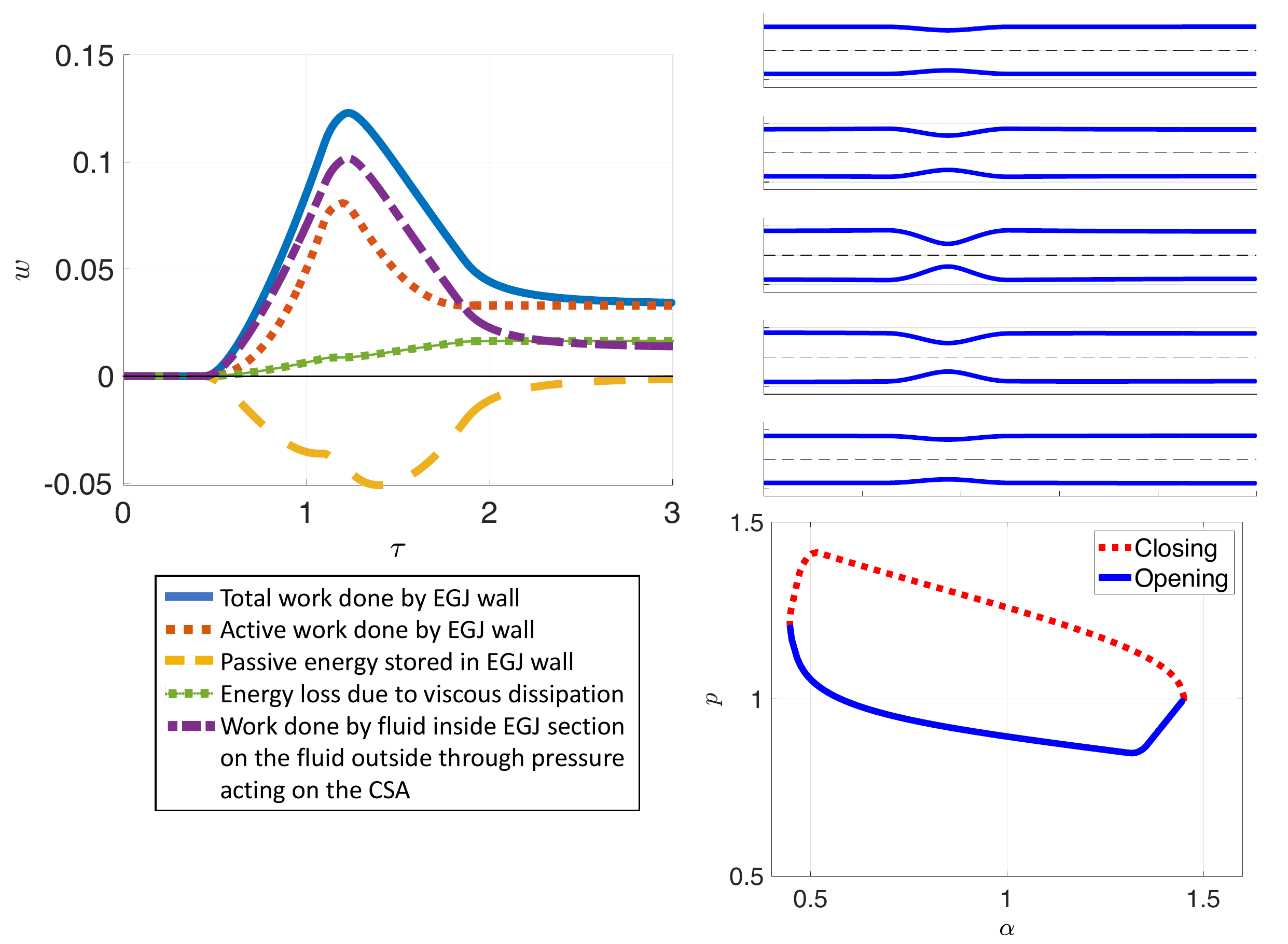}}}
    \caption{Simulation results of a single contraction cycle with muscle activity defined in figure \ref{fig:activationSqueez}. The left plot shows the work components from equation (\ref{eq:work_balance}) evaluated at the sphincter region and the total work done by the sphincter wall on the fluid (LHS of equation (\ref{eq:work_balance_NonDim_no_split}). The plot at the bottom right presents the corresponding pressure-CSA loop as recorded at the sphincter location. The top right plot displays the tube shape at five consecutive instants, ordered chronologically.}
    \label{fig:simulationSqueezingResults}
\end{figure*}

As long as tone relaxation is the only component resulting in the opening of the sphincter, a NSL emerges, independent of the order of operations. For instance, figure \ref{fig:simulationSqueezingResults} presents the results of a simulation with activation function defined in figure \ref{fig:activationSqueez}, where a relaxed sphincter contracts and then relaxes back. As the figure shows, this simulation obtained a NSL (bottom right plot). The resulting tube shape at five consecutive instances are displayed on the top right of the figure. In this example, $\beta=5000$, $\psi=100$, and the sphincter's width is $25\%$ of the tube length ($\text{w}_{\text{s}}=0.25$). We choose to exhibit a scenario with a larger $\beta$ value to show the effect of fluid viscosity on the result.


The plot on the left in figure \ref{fig:simulationSqueezingResults} presents the work components in equation (\ref{eq:work_balance}) and the total work term on the left hand side of equation (\ref{eq:work_balance_NonDim_no_split}), normalized by $\psi$ ($w=\text{Work}/\psi$), as a function of time. This plot gives us an insight into the underlying mechanism relating pressure and CSA in this case. As the figure shows, the total work done by the sphincter on the fluid (left hand side of equation (\ref{eq:work_balance_NonDim_no_split})) has the opposite trend to the curve in figure \ref{fig:simulationToneRelaxResults1}. This is due to the fact that in this simulation, the sphincter wall first contracts, doing positive work on the fluid (total work curve increases) and then relaxes, doing negative work on the fluid (total work curve decreases). This trend, although opposite, still explains the same mechanism which results in a NSL. Initially, the bag in the tube is pressurized while the sphincter is mostly open. Hence, once the sphincter starts to contract, decreasing CSA, the pressure in the bag increases, resulting in a negative slope curve between pressure and CSA. This is the pattern that is observed in AS experimental set up by \cite{Zifan2019}.

Similar to the results in figure \ref{fig:simulationToneRelaxResults1}, the net total work done by the wall on the fluid in figure \ref{fig:simulationSqueezingResults} is positive, indicating that the wall does more work on the fluid than the fluid (through pressure) does on the wall throughout this opening and closing cycle. 

It is also interesting to look at the passive and active decomposition of the total work, as the trends are slightly different than the ones discussed in figure \ref{fig:simulationToneRelaxResults}. In the simulation discussed in figure \ref{fig:simulationSqueezingResults}, the active work done by the sphincter wall on the fluid has a similar trend to the one previously described. The active work done by the sphincter wall on the fluid increases as the wall contracts, because the direction of the force exerted by the wall on the fluid is pointed in the same direction as the wall motion. When the sphincter reaches maximum contraction, it reaches its minimum CSA and there is no change in active work. Once the sphincter starts to relax back,  CSA increases, and the active work done by the sphincter wall on the fluid begins to decrease since the force exerted by the wall on the fluid is opposite to the direction of wall motion. The elastic energy stored in the tube wall in this simulation has an opposite pattern than the one described before (figure \ref{fig:simulationToneRelaxResults}). Since the CSA decreases as a direct result of squeezing, passive work, which captures this change, is negative.

Notice that the area between the opening and closing curves in figure \ref{fig:simulationSqueezingResults} is much greater than the one in figure \ref{fig:simulationToneRelaxResults1}. As elaborated upon in \cite{Elisha2021Loop}, the loop area indicates that there is some energy that is being gained or lost by the system. If the closing curve is above the opening curve, as in the two cases discussed above, energy is being lost, meaning that the sphincter does most of the work in the opening and closing cycle (as opposed to pressure from the fluid). Thus, the larger the area between the two curves, the more energy the sphincter tone needs to expend. The sphincter wall applies more work in the example presented in figure \ref{fig:simulationSqueezingResults} than the one in figure \ref{fig:simulationToneRelaxResults1}. This difference is mostly due to the value of $\beta$ (equivalent to increasing fluid viscosity). Increasing fluid viscosity (by increasing $\beta$) implies increasing flow resistance in this case \cite{Elisha2021}, which requires more work from the wall in order to contract. The effect of viscosity is further discussed in \cite{Elisha2021Loop}.

\begin{figure*}[!htb]
    \centering{{\includegraphics[trim=0 120 0 120 ,clip,width=0.9\textwidth]{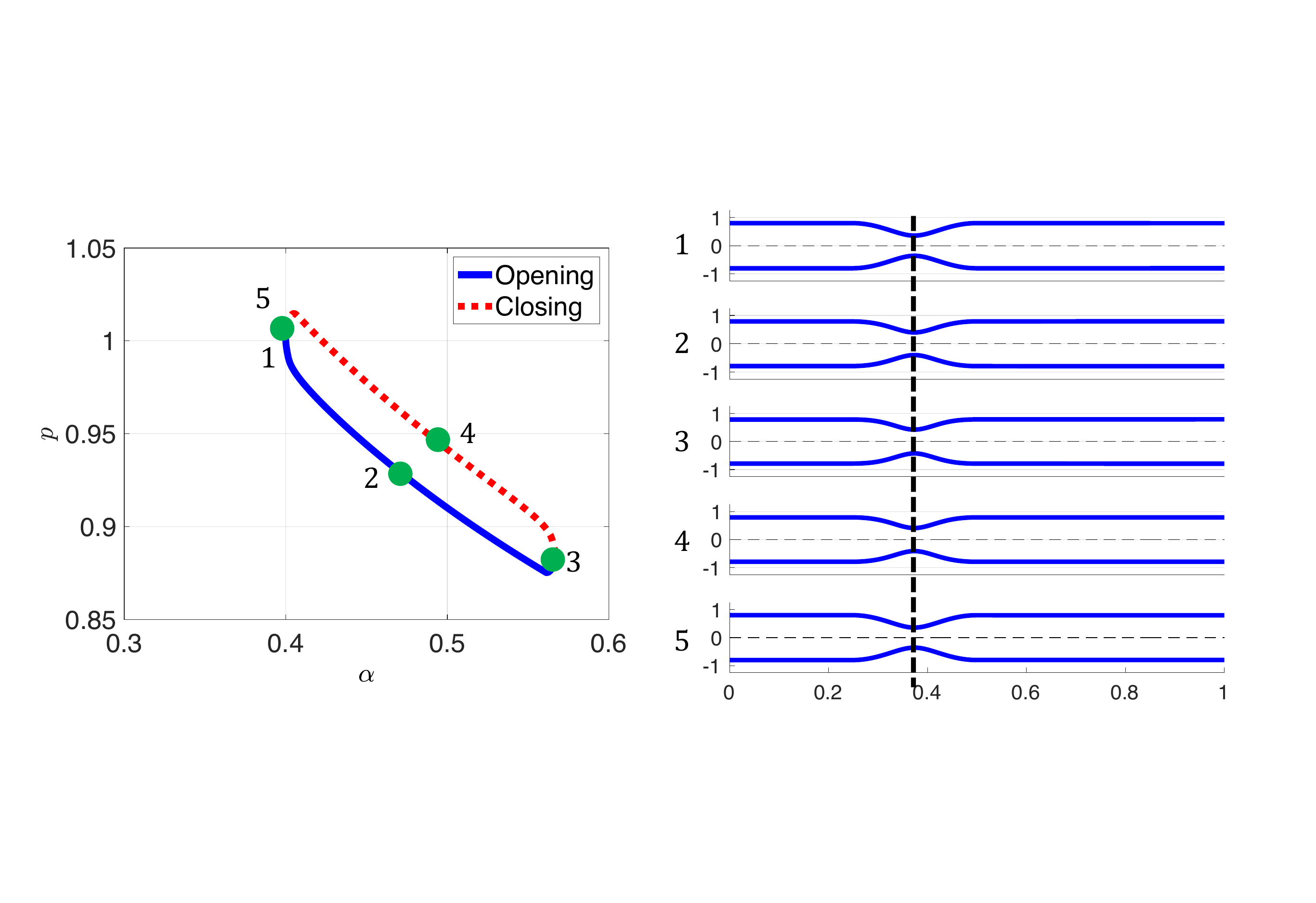}}}
    \caption{Simulation results of a single contraction cycle with constant reference CSA but varying external pressure, $p_o$ (figure \ref{fig:PoFunct}). Pressure-CSA loop at the sphincter location (left) and the tube shapes at five consecutive instances (right).}
    \label{fig:uesSimResults}
\end{figure*}

Note that a NSL is also observed from simulations where $\theta$ does not vary in time but $p_o=p_o(\chi,\tau)$, as in the UES. In these simulations, equation (\ref{eqn:tube_law_delta}) is used for the constitutive pressure relation, where $p_o(\chi,\tau)$ is illustrated in figure \ref{fig:PoFunct}. Figure \ref{fig:uesSimResults} presents the results obtained by a simulation of this set up, with input parameters $\beta=100$, $\psi=100$, and $\text{w}_{\text{s}}=0.25$. The left plot displays the corresponding pressure-CSA loop at the sphincter location (exhibiting a NSL). The five points marked on the loop correspond to five time instants in the contraction cycle. The tube shapes at these five instants are plotted on the right. These results emphasize once again that when the sphincter opens due to mechanism (i), a NSL emerges. Furthermore, it strengthens the argument that the mechanical imprint of the pressure-CSA loop does not change whether the opening is due to neurogenic mediated relaxation or neurally controlled external pulling. The mechanical picture of both scenarios is equivalent and therefore labeled as mechanism (i). Notice that the change in CSA is smaller than the one obtained from simulations where the sphincter opens by muscle relaxation (change reference CSA). This is simply because changing external pressure in the simulation context is not as effective as varying reference CSA.

\subsection{Positive Slope Pressure-Area Loop} \label{PosSlope}

\begin{figure*}[!htb]
    \centering{{\includegraphics[trim=0 100 0 95 ,clip,width=0.9\textwidth]{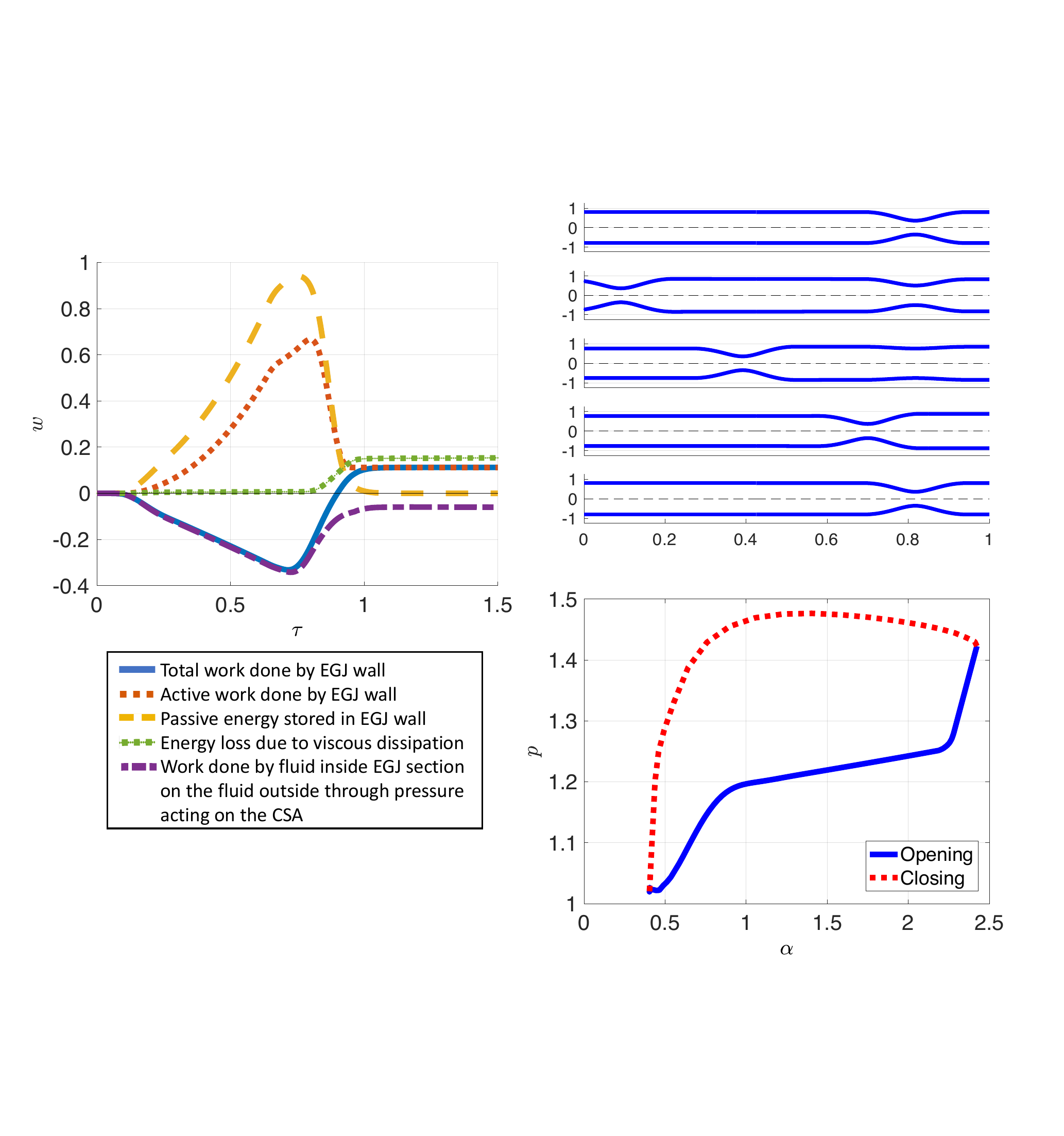}}}
    \caption{Simulation results of a single contraction cycle with muscle activity defined by a neurally controlled tone relaxation and a traveling peristalsis (figure \ref{fig:activationEGJ}). The left plot shows the work components from equation (\ref{eq:work_balance}) evaluated at the sphincter region and the total work done by the sphincter wall on the fluid (LHS of equation (\ref{eq:work_balance_NonDim_no_split}). The plot at the bottom right presents the corresponding pressure-CSA loop as recorded at the sphincter location. The top right plot displays the tube shape at five consecutive instants, ordered chronologically.}
    \label{fig:simulationEGJResults}
\end{figure*}

Figure \ref{fig:simulationEGJResults} presents the results obtained by a simulation of a single contraction cycle with the activation function presented in figure \ref{fig:activationEGJ}, where the muscle activity is defined by both a sphincter tone and a traveling peristalsis. Hence, the opening of the sphincter is a result of both neurogenic mediated relaxation and mechanical distention. In this simulation, $\beta=100$, $\psi=100$, and the width of both the traveling wave and the sphincter is $25\%$ of the tube length ($\text{w}_{\text{s}}=0.25$). The resulting tube shapes at five consecutive instances are presented on the top right of the figure. The bottom right plot shows the corresponding pressure-CSA loop at the sphincter location. As the figure shows, a PSL emerges. 

This remark suggests that when the sphincter opening pattern involves both mechanical distention and neurogenic mediated opening, as in the case of the LES, a PSL is obtained. The reason for that lies in the way in which the activation function dictates the pressure in the tube. In the previous two cases (figures \ref{fig:simulationToneRelaxResults1} and \ref{fig:simulationSqueezingResults}), when the sphincter tone starts to relax, the sphincter's CSA increases, and consequently the pressure decreases. In this case (figure \ref{fig:simulationEGJResults}), a similar process takes place, however, as sphincter relaxation starts, a peristaltic contraction begins traveling down the tube length, which increases pressure that mechanically helps in opening the sphincter. 

The left plot in figure \ref{fig:simulationEGJResults} presents the work components in equation (\ref{eq:work_balance}) and the total work term on the left hand side of equation (\ref{eq:work_balance_NonDim_no_split}), normalized by $\psi$ ($w=\text{Work}/\psi$), as a function of time. As the figure shows, the individual curves follow similar patterns to the ones appear in figure \ref{fig:simulationToneRelaxResults}. This is simply because the CSA of the sphincter changes in the same way (i.e., contracted $\rightarrow$ relaxed $\rightarrow$ contracted). The main difference between the two plots is in the passive work curve, which is a lot larger than the active work curve in figure \ref{fig:simulationEGJResults} as opposed to figure \ref{fig:simulationToneRelaxResults}. This trend results in the two pressure-CSA patterns. The passive work in figure \ref{fig:simulationEGJResults} is much larger than the active work because it increases as a result of mechanical distention created by the peristaltic contraction wave in addition to tone relaxation.

As revealed in section \ref{LESLoops}, clinical pressure-CSA loops at the LES display two types. In our previous LES loop study, we exposed the different mechanisms and properties that result in the different loops \cite{Elisha2021Loop}. PDL is associated with high tube stiffness, low fluid viscosity, and slow LES relaxation, whereas TDL is associated with low tube stiffness, high fluid viscosity, and fast LES relaxation \cite{Elisha2021Loop}. Relaxation speed means how long it takes the contracted sphincter, once the peristaltic wave has entered the domain, to go from fully contracted to fully relaxed. The relaxation speed was implemented to the model through the activation function. Since the fluid in our clinical FLIP study is always saline, fluid viscosity is the same for all clinical cases. Moreover, esophagus stiffness is high and approximately equal for all control cases. Thus, it was concluded that the appearance of two loop types in clinical data is a result of neurally controlled activation function and not material or fluid properties \cite{Elisha2021Loop}. This conclusion aligns with the conclusions presented in the above study, where muscle activation pattern dictates the pressure-CSA relation.

\section{Discussion} \label{Discussion}

Knowing what is the controlling mechanism that dictates the pressure-CSA loop types at the different sphincters has several useful applications. First, we can use this to hypothesize how the pressure-CSA loops will look for sphincters where data is not available. For instance, limited work has been done to study the PS using FLIP and we did not find any FLIP studies capturing the opening and closing trends of the PS during a contraction cycle. Knowing that the PS opening and closing cycle during the inter-digestive state is controlled by peristalsis and tone relaxation, we can hypothesize that PS pressure-CSA plots throughout single cycles will display a positive slope. The loop type for each sphincter, based on the information revealed in this study, is hypothesized and listed in table \ref{table:loopTypeTable}. Second, we can use information from the pressure-CSA loops to identify normal and abnormal phenotypes for the different sphincters. Lastly, calculating the work done by the sphincter can help in creating a physiomarker. This can be used as a diagnostic tool to evaluate pathophysiologies of sphincters.


\begin{table}[h]
\centering
   \caption{Human sphincters and the hypothesized resulting pressure-CSA loops obtained by FLIP experiments}
    \label{crouch}
    \begin{tabular}{  p{0.3\textwidth}  p{0.2\textwidth}  }
        \toprule
\textbf{Sphincter}      
& \textbf{Description} \\\midrule
Upper esophageal sphincter
& Negative slope \\\hline
Lower esophageal sphincter       
& Positive slope \\\hline
Pyloric sphincter       
& Positive slope\\\hline
Sphincter of Oddi&
Negative slope\\\hline
Internal anal sphincter&
Negative slope\\\hline
External anal sphincter&
Negative slope\\\hline 
Urethral sphincter&
Negative slope\\
        \bottomrule
    \end{tabular}
    \label{table:loopTypeTable}
\end{table}

Although both peristalsis and tone relaxation are present at the SoO, clinical data reveals that the SoO pressure-CSA relation depicts negative slope characteristics \cite{kunwald2010new}. Recall that the loop type depends on the amount of pressurization and the timing of the opening/relaxation of the sphincter itself. Thus, this indicates that relaxation is dominating the opening of the SoO and the pressurization has little effect.

Note that the loop types and their underlying mechanisms discussed in this report only hold for FLIP manometry measurements of a bag with fluid at constant volume. Other measuring devices which relate pressure and CSA may result in different loop patterns. For example, high-resolution manometry has been used by many to measure pressure and CSA at different sphincters \cite{Omari2016,Omari2015,Lin2014,Carlson2015high,Kirby2015,Desipio2007,Wasenda2018,Vitton2013,Carrington2014,Jones2007}. Similar to the FLIP, it can be used to relate changes in CSA and the corresponding changes in pressure, recorded at the same instance and location \cite{Kahrilas2015,Carlson2015high,Pandolfino2006}. However, few studies have plotted the pressure-CSA hysteresis and attempted to explain the observations. \cite{Omari2015} and \cite{Omari2016} plotted the pressure-CSA relation at the UES using high-resolution manometry and classified possible mechanical states of the UES muscle, which take place during the opening and closing cycle. Their results reveal an L-shaped pressure-CSA relation. These observations are not reproduced nor discussed in this work, but should be the subject for further investigation.

Lastly, the experimental study presented in this work has a significant limitation. The FLIP bag's compliance limit is 380 $\text{mm}^2$, which is passed in almost all cases. However, this fact does not alter the conclusions presented in this report, as the qualitative trend and pattern of the LES and peristalsis muscle activity is still observed and conserved.

\section{Conclusion} \label{Conclusion}


This study aimed to provide a consolidated approach one can take to investigate the function of human sphincters. Borrowed from cardiovascular literature, this report presents a systematic analysis of the opening and closing pattern of human sphincters through looking at their pressure-CSA hysteresis during a single contraction cycle measured by FLIP or FLIP-like devices. Based on literature review and our experimental investigation of the LES, we identified two shapes of possible pressure-CSA hysteresis. We hypothesized and then showed using simulations that the presence (or absence) of two mechanical modes dictates the resulting loop type. The study revealed that when the opening of a sphincter is controlled by the sole activity of the sphincter muscle relaxing, a NSL emerges. When the opening of the contracted sphincter is a function of both neurogenic mediated relaxation and mechanical distention (through peristalsis), a PSL is obtained. In addition, through examining the resulting pressure-CSA loops at the LES, which resulted in PSL, we identified two major loop types which was discussed in greater detail in our previous work \cite{Elisha2021Loop}.


\bibliographystyle{asmejour}   
\bibliography{SphincterManuscript_bib} 

\end{document}